\documentclass{elsart}

\usepackage{amsmath,amssymb}
\usepackage{graphicx}
\usepackage{url}
\usepackage{cite}

\newlength\figwidth
\newlength\subfigskip
\begin{document}

\begin{frontmatter}

\title{Breaking a chaos-based secure communication scheme designed by an improved modulation method}
\thanks{This paper has been published by in \textit{Chaos, Solitons \& Fractals}, vol. 25, no. 1, pp.
109-120, 2005, and has been available online at
\url{http://dx.doi.org/10.1016/j.chaos.2004.09.077}.}

\author[China]{Shujun Li\corauthref{corr}},
\author[Spain]{Gonzalo \'{A}lvarez} and
\author[China]{Guanrong Chen}
\address[China]{Department of Electronic Engineering, City
University of Hong Kong, 83 Tat Chee Avenue, Kowloon Tong, Hong
Kong SAR, China}
\address[Spain]{Instituto de F\'{\i}sica
Aplicada, Consejo Superior de Investigaciones Cient\'{\i}ficas,
Serrano 144---28006 Madrid, Spain}

\corauth[corr]{The corresponding author, email address:
\url{hooklee@mail.com}, personal web site:
\url{http://www.hooklee.com}.}

\begin{abstract}
Recently Bu and Wang \cite{BuWang:CSF2004} proposed a simple
modulation method aiming to improve the security of chaos-based
secure communications against return-map-based attacks. Soon this
modulation method was independently cryptanalyzed by Chee et al.
\cite{CheeXuBishop:CSF2004}, Wu et al. \cite{WuHuZhang:CSF2004},
and \'{A}lvarez et al. \cite{Alvarez:CSF2004} via different
attacks. As an enhancement to the Bu-Wang method, an improving
scheme was suggested by Wu et al. \cite{WuHuZhang:CSF2004} by
removing the relationship between the modulating function and the
zero-points. The present paper points out that the improved scheme
proposed in \cite{WuHuZhang:CSF2004} is still insecure against a
new attack. Compared with the existing attacks, the proposed
attack is more powerful and can also break the original Bu-Wang
scheme. Furthermore, it is pointed out that the security of the
scheme proposed in \cite{WuHuZhang:CSF2004} is not so satisfactory
from a pure cryptographical point of view. The synchronization
performance of this class of modulation-based schemes is also
discussed.
\end{abstract}

\end{frontmatter}

\section{Introduction}

Since the early 1990s, the use of chaotic systems in cryptography
has been extensively investigated. There are two main classes of
chaotic cryptosystems: analog \cite{Alvarez:Survey:ICCST99,
Yang:Survey:IJCC2004} and digital
\cite{ShujunLi:Dissertation2003}. Most analog chaotic
cryptosystems are secure communication systems based on
synchronization of the sender and the receiver chaotic systems
\cite{Pecora:CS:PRL90}, where a signal is transmitted over a
public channel from the sender to the receiver, and decryption of
the plain-signal is realized via chaos synchronization at the
receiver end. According to the encryption structures, most analog
chaos-based secure communication systems can be classified into
four categories: chaotic masking \cite{Kocarev:CM:IJBC92,
Feki:CM:PLA99}, chaotic switching or chaotic shift keying (CSK)
\cite{Parlitz:CSK:LIJBC92, Dedieu:CSK:IEEETCASII93}, chaotic
modulation \cite{Yang:CPM:IEEETCASI96, Parlitz:CDM:PRE96}, and
chaotic inverse system \cite{Feldmann:ISA:IJCTA96}. Meanwhile,
some cryptanalysis work has also been developed, and many
different attacks have been proposed to break different types of
chaos-based secure communication systems: return-map-based attacks
\cite{Perez:ReturnMapCryptanalysis:PRL95,
Zhou:ExtractChaoticSignal:PLA97,
Yang:ReturnMapCryptanalysis:PLA98}, spectral analysis attacks
\cite{Yang:SpectralCryptanalysis:PLA98, Alvarez-Li2004a},
generalized-synchronization-based attacks
\cite{Yang:GSCryptanalysis:IEEETCASI97,
Alvarez:BreakingCPM:CSF2004}, short-time-period-based attacks
\cite{Yang:STZCR:IJCTA95, Alvarez-Li2004b}, prediction-based
attacks \cite{Short:ChaoticSignalExtraction:IJBC97,
ZhouLai:ChaoticCryptanalysis:PRE99b},
parameter-identification-based attacks
\cite{Kocarev:BreakingParameters:IJBC96,
TaoDu:ChaoticCrytpanalysis:IJBC2003a,
Vaidya:DecodingChaoticSuperkey:CSF2003,Alvarez:BreakingPhase:Chaos04},
and so on.

Since many early chaos-based secure communication systems are
found insecure against various attacks, how to design robust
chaos-based secure communication systems against the existing
attacks is always a real challenge. The following three types of
countermeasures have been proposed in the literature: 1) using
more complex dynamical systems, such as hyperchaotic systems or
multiple cascaded heterogeneous chaotic systems
\cite{Murali:HeterogeneousChaosCryptography:PLA2000}; 2)
introducing traditional ciphers into chaotic cryptosystems
\cite{Yang-Wu-Chua:ChaoticCryptography:IEEETCASI1997}; 3)
introducing an impulsive (also named sporadic) driving signal
instead of a continuous signal to realize modulation and
synchronization \cite{Yang-Chua:ImpulsiveChaos:IJBC1997}. The
first countermeasure was found lately remaining insecure against
some specific attacks \cite{Short:UnmaskingHyperchaos:PRE98,
ZhouLai:ChaoticCryptanalysis:PRE99b,
Huang:UnmaskingChaosWavlet:IJBC2001,
TaoDu:ChaoticCrytpanalysis:IJBC2003a}, and some security defects
of the second have also been reported
\cite{Short:ChaoticCrptanalysis:IEEETCASI2001}.

In \cite{BuWang:CSF2004}, a simple modulation method was proposed
by Bu and Wang to enhance the security of some chaos-based secure
communications against return-map-based attacks. A periodic signal
is used to modulate the transmitted signal in order to effectively
blur the reconstructed return map. Although this modulation method
can frustrate return-map-based attacks, it was soon broken
independently in \cite{CheeXuBishop:CSF2004, WuHuZhang:CSF2004,
Alvarez:CSF2004} with other different attacks, some of which rely
on the embedded periodicity of zero-crossing points of the
modulating signal. To further improve the security of the original
Bu-Wang modulation method, a modified modulating signal was then
suggested in \cite{WuHuZhang:CSF2004} to cancel the embedding
regular occurrence of zero-crossing points in the transmitted
signal.

This paper points out that the modified modulation method proposed
in \cite{WuHuZhang:CSF2004} is still insecure against a new attack
that cab restore the return map. The secret parameters of the
modulating signal can be approximately identified via the proposed
attack. Compared with other attacks, the new attack is more
powerful which can also break the original Bu-Wang modulation
scheme. Furthermore, it will be pointed out that from a
cryptographical point of view the modulation method itself is not
so satisfactory in enhancing the security of the chaotic
cryptosystems even when the modulating signal is secure enough.

The rest of this paper is organized as follows. In Sec.
\ref{section:modulating}, the modulation method and related
attacks are briefly introduced. Section
\ref{section:cryptanalysis} shows how the new attack breaks the
scheme based on the modified modulation method. Some general
discussions on practical performance of this class of
modulation-based methods under study are given in Sec.
\ref{section:discussions}. The last section concludes the paper.

\section{The modulation method and related attacks}
\label{section:modulating}

To introduce the modulation-based method, the Lorenz system is
used to construct a secure communication system. The sender system
is
\begin{eqnarray}
\dot{x}_1 & = & \sigma(x_2-x_1),\nonumber\\
\dot{x}_2 & = & rx_1-x_2-x_1x_3,\\
\dot{x}_3 & = & x_1x_2-bx_3\nonumber
\end{eqnarray}
and the receiver system is
\begin{eqnarray}
\dot{y}_1 & = & \sigma(y_2-y_1)+c(s(\mathbf{x},t)-s(\mathbf{y},t)),\nonumber\\
\dot{y}_2 & = & ry_1-y_2-y_1y_3,\\
\dot{y}_3 & = & y_1y_2-by_3,\nonumber
\end{eqnarray}
where $\sigma,b,r$ are parameters of the system,
$\mathbf{x}=(x_1,x_2,x_3)$ and $\mathbf{y}=(y_1,y_2,y_3)$ are
state variables, $c$ denotes the coupling strength, and
$s(\mathbf{x},t)=g(t)x_1(t)$ is the ciphertext transmitted over
the public channel for information carrying and chaos
synchronization. Accordingly, $s(\mathbf{y},t)=g(t)y_1(t)$. In
\cite{BuWang:CSF2004}, $g(t)$ is selected as the product of a
cosine signal and another system variable: $g(t)=A\cos(\omega
t+\phi_0)x_3(t)$. When the above system runs in a chaotic masking
configuration, $s(\mathbf{x},t)=g(t)x_1(t)+i(t)$, where $i(t)$ is
the plain-message signal.

In the above secure communication system, the secret key consists
of the system parameters of the Lorenz system $(\sigma,b,r)$ and
the parameters of the modulating signal $(\omega,\phi_0)$. Note
that $A$ should not be used as part of the key, since it does not
influence the shape of the return map (just change the size). To
facilitate the following discussion, without loss of generality,
the parameters are fixed as the default parameters used in
\cite{BuWang:CSF2004}: $\sigma=16,r=45.6,b=4.0$,
$A=0.5,\omega=1.5,\phi_0=0$. In the chaotic masking configuration,
the plain-message signal is assumed to be $i(t)=0.1\sin(t)$; in
the chaotic switching configuration, the value of $b$ switches
between $b_0=4.0$ and $b_1=4.4$.

The purpose of using $g(t)$ is to frustrate return-map-based
attacks, proposed previously in
\cite{Perez:ReturnMapCryptanalysis:PRL95}. In common chaotic
masking systems, $s(\mathbf{x},t)=x_1(t)+i(t)$, where $i(t)$ is
the plain-message signal whose energy is much smaller than that of
$x_1(t)$; and in common chaotic switching systems,
$s(\mathbf{x},t)=x_1(t)$. Following
\cite{Perez:ReturnMapCryptanalysis:PRL95}, two return maps, $A_n$
vs $B_n$ and $(-C_n)$ vs $(-D_n)$, are defined as follows:
$A_n=\frac{X_n+Y_n}{2}$, $B_n=X_n-Y_n$,
$C_n=\frac{X_{n+1}+Y_n}{2}$, $D_n=Y_n-X_{n+1}$, where $X_n$ and
$Y_m$ are the $n$-th (local) maxima and the $m$-th (local) minima
of the transmitted signal, respectively. Since the $A_n$ vs $B_n$
map is identical to the $(-C_n)$ vs $(-D_n)$ map, in this paper
only the former will be considered. Once an attacker gets the
$A_n$ vs $B_n$ return map, he can use the attack method proposed
in \cite{Perez:ReturnMapCryptanalysis:PRL95} to break both the
chaotic masking and the chaotic switching systems.

\begin{figure}
\centering
\begin{minipage}[t]{\figwidth}
\centering
\includegraphics[width=\textwidth]{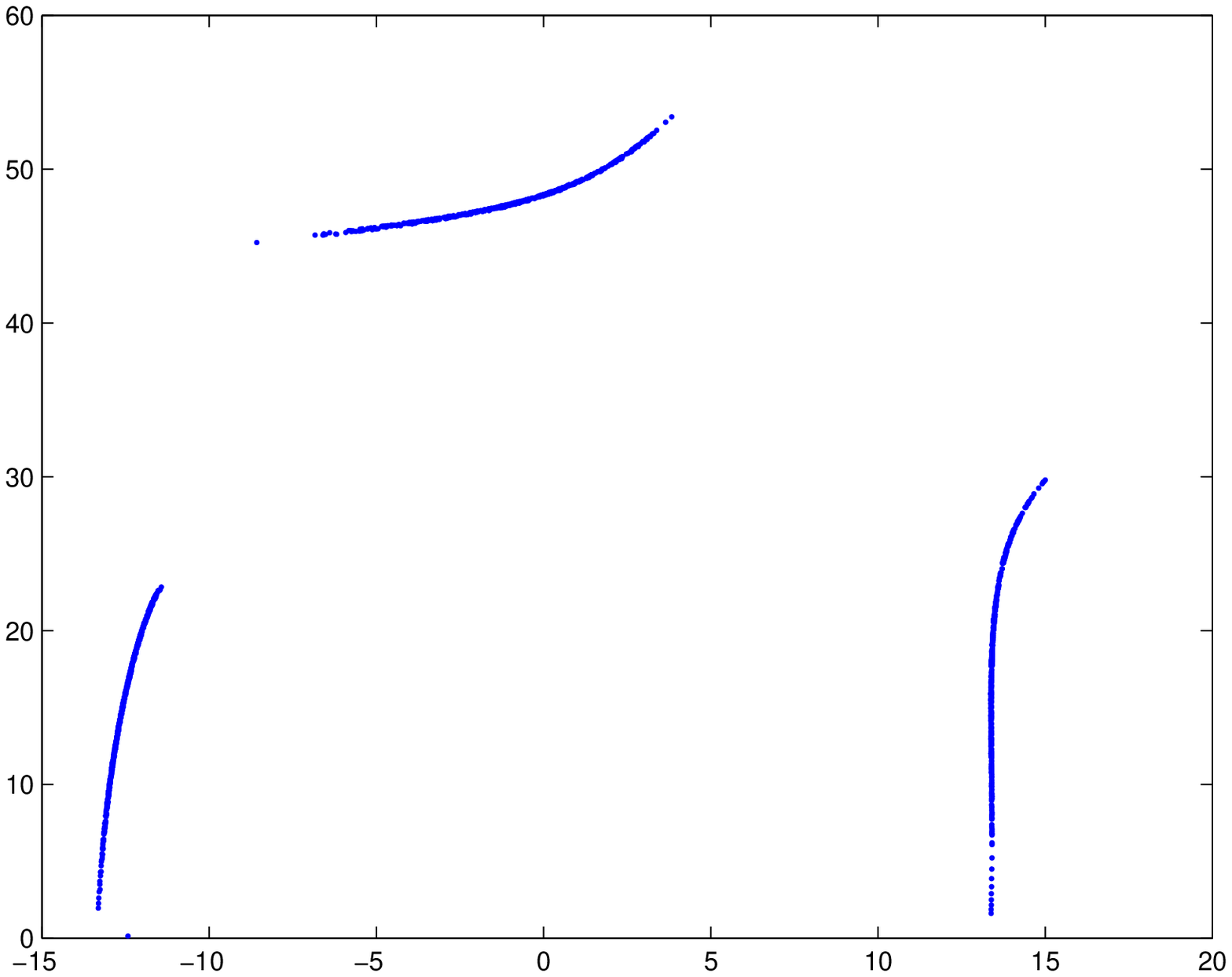}
a) chaotic carrier: $x_1(t)$
\end{minipage}
\begin{minipage}[t]{\figwidth}
\centering
\includegraphics[width=\textwidth]{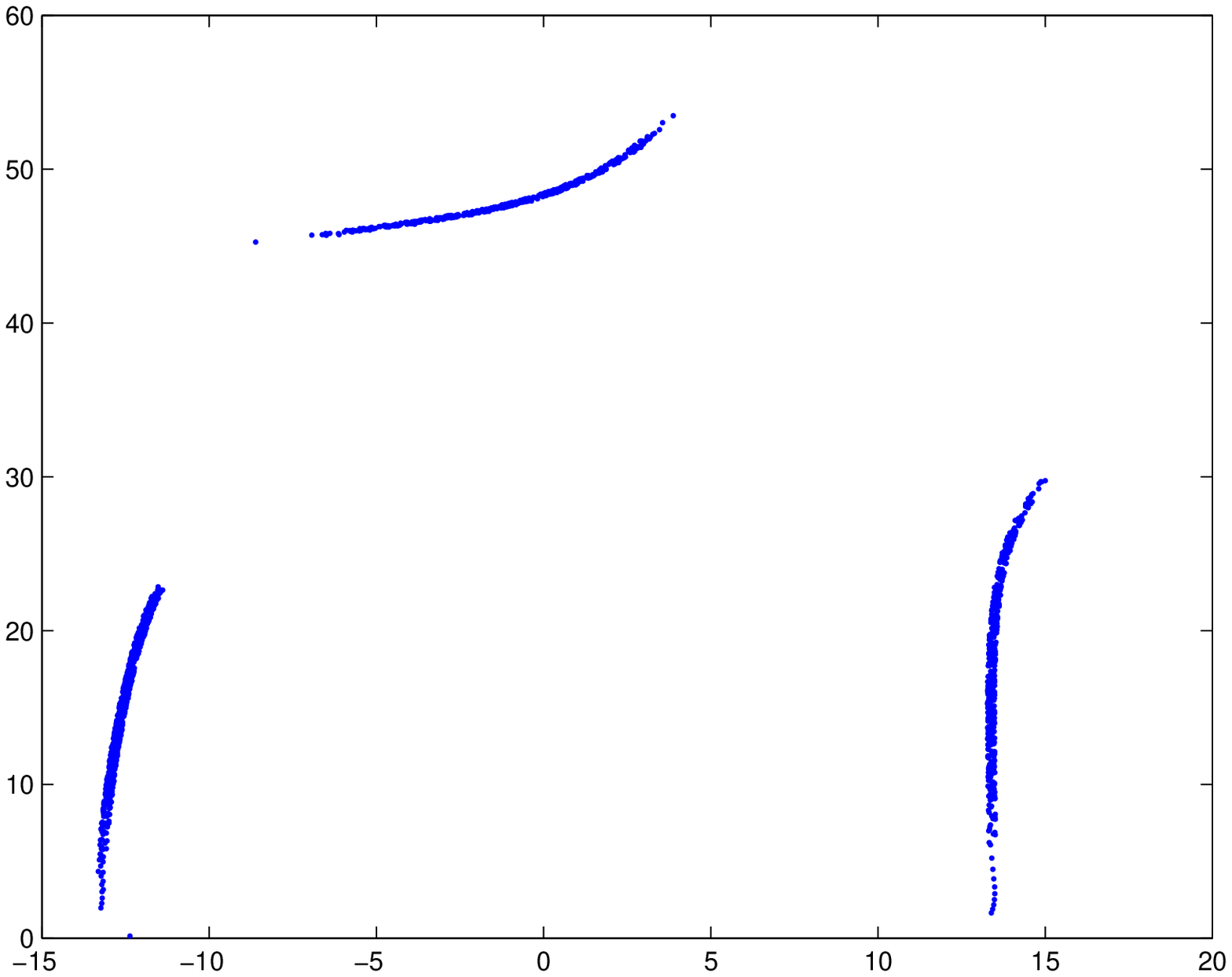}
b) chaotic masking: $x_1(t)+i(t)$
\end{minipage}
\begin{minipage}[t]{\figwidth}
\centering
\includegraphics[width=\textwidth]{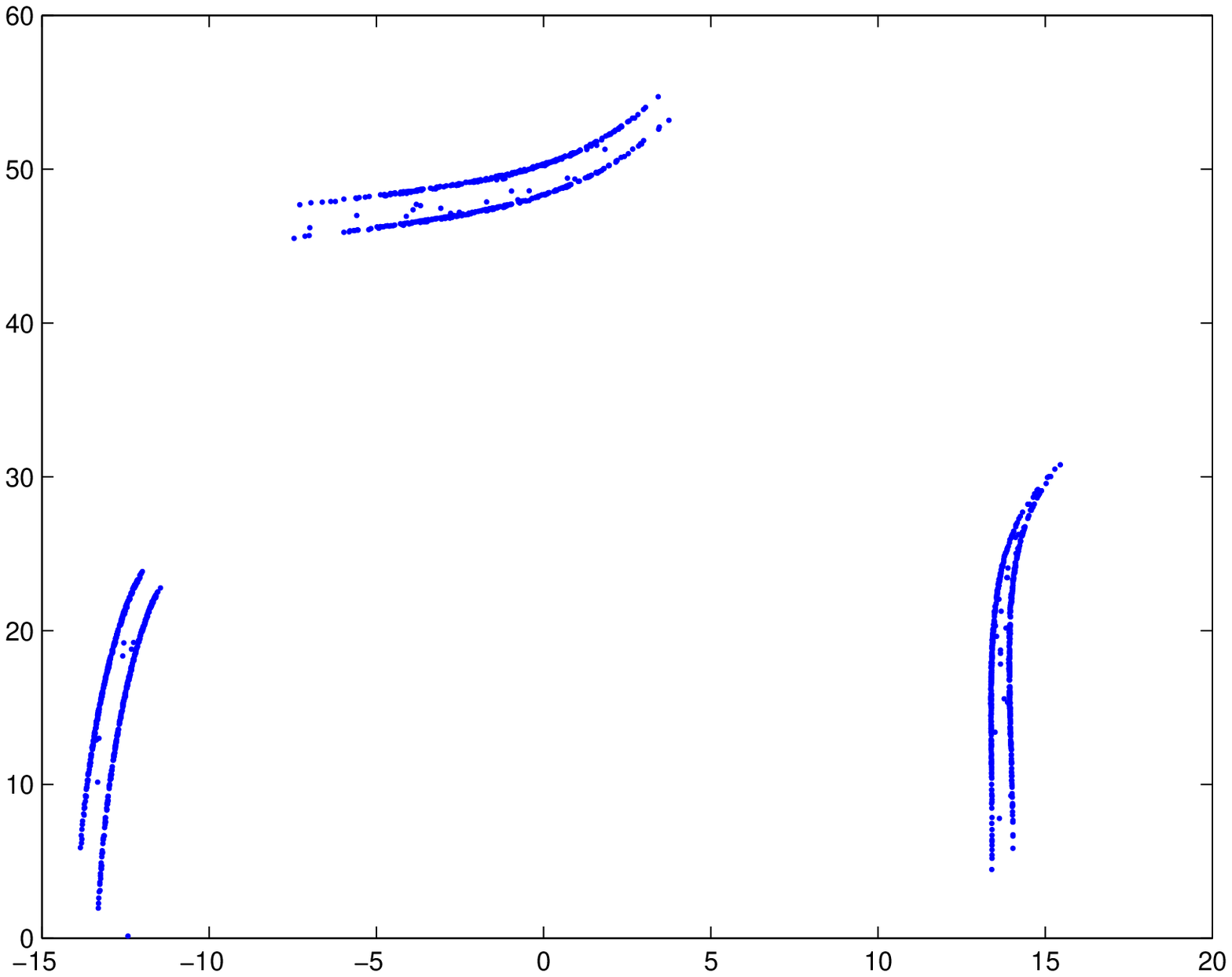}
c) chaotic switching: $x_1(t)$
\end{minipage}\\[\subfigskip]
\begin{minipage}[t]{\figwidth}
\centering
\includegraphics[width=\textwidth]{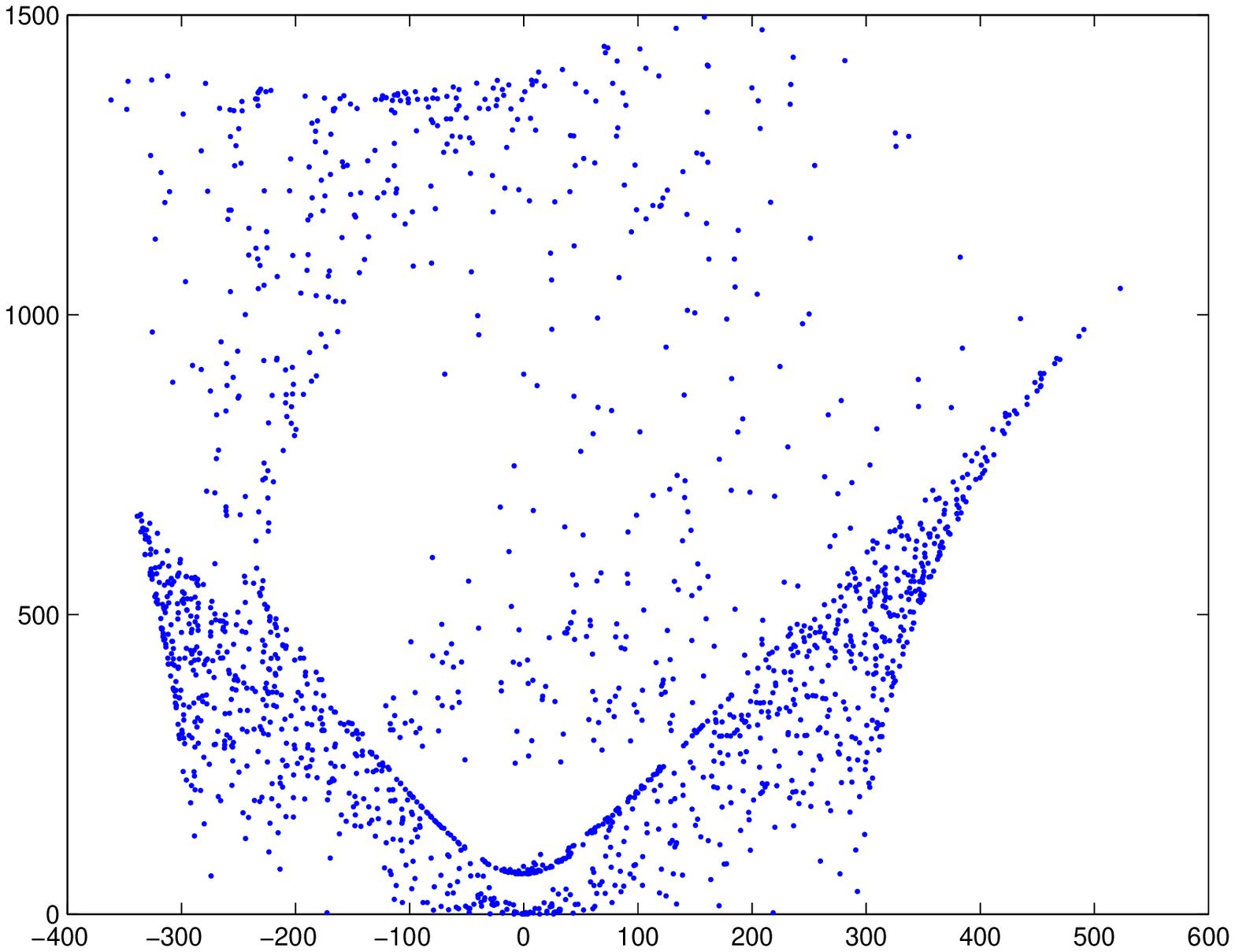}
d) chaotic carrier: $g(t)x_1(t)$
\end{minipage}
\begin{minipage}[t]{\figwidth}
\centering
\includegraphics[width=\textwidth]{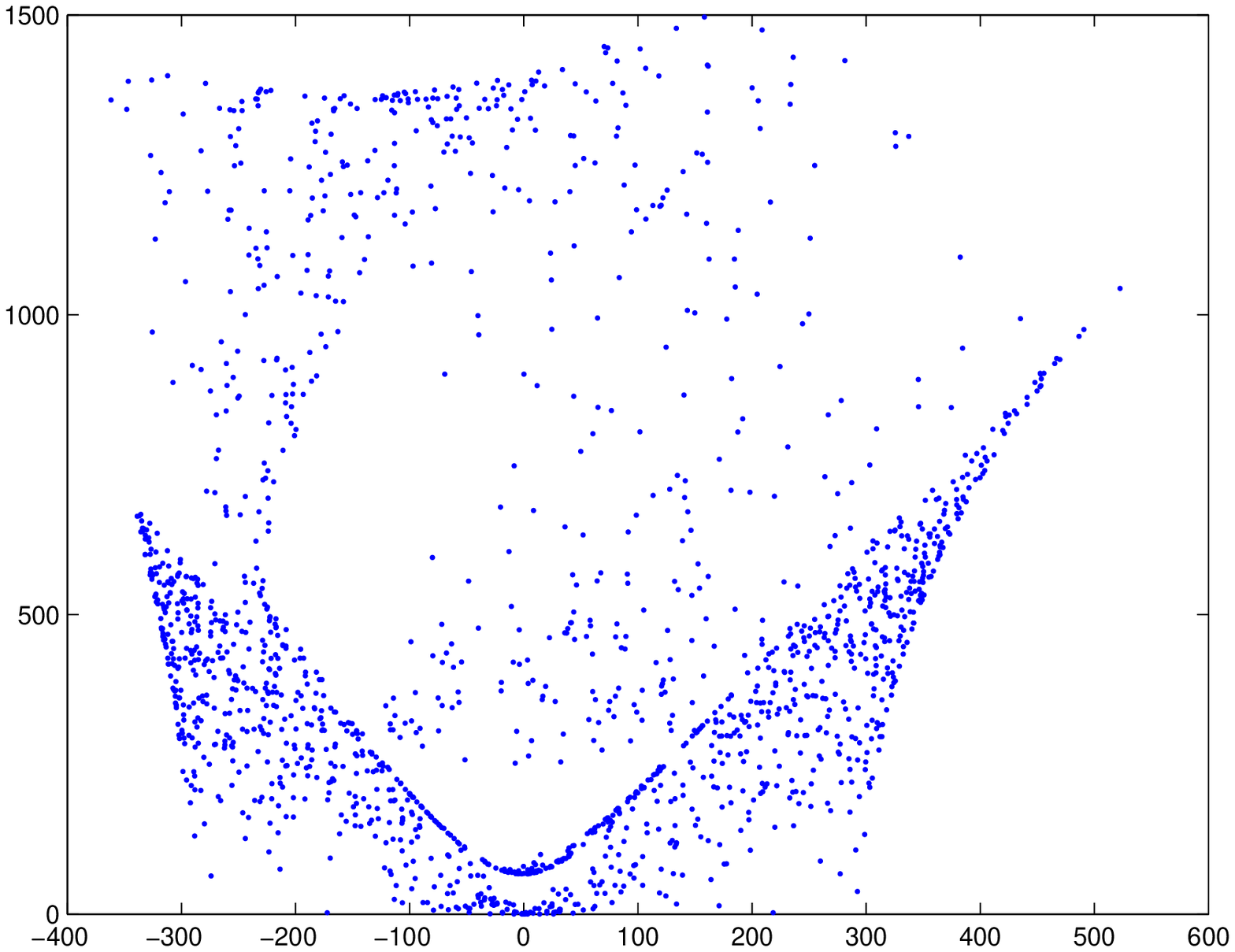}
e) chaotic masking: $g(t)x_1(t)+i(t)$
\end{minipage}
\begin{minipage}[t]{\figwidth}
\centering
\includegraphics[width=\textwidth]{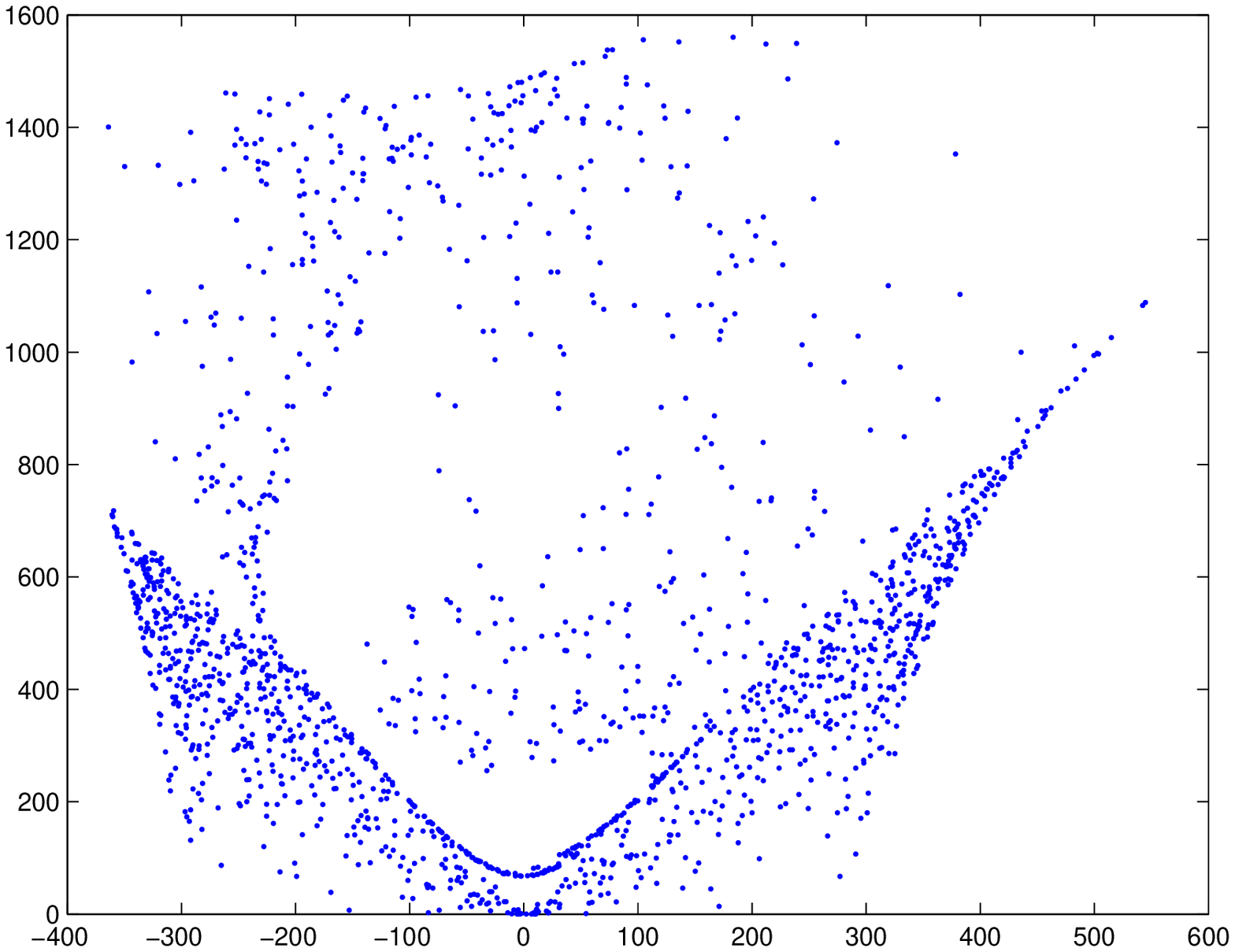}
f) chaotic switching: $g(t)x_1(t)$
\end{minipage}
\caption{Return maps ($A_n$ vs $B_n$, the same thereinafter)
reconstructed under different conditions.}
\label{figure:ReturnMap0}
\end{figure}

In Fig. \ref{figure:ReturnMap0}, some return maps reconstructed
under different conditions are shown, from which it seems that the
modulating function $g(t)=A\cos(\omega t+\phi_0)x_3(t)$ can
effectively blur the return maps and so frustrate the attackers.
However, as pointed out in \cite{CheeXuBishop:CSF2004,
WuHuZhang:CSF2004}, the modulating signal will introduce regular
zero-crossing points in the transmitted signal $g(t)$, which makes
it possible to distinguish the value of $\omega$ via spectral
analysis \cite{CheeXuBishop:CSF2004} or autocorrelation analysis
\cite{WuHuZhang:CSF2004}. Also, it was shown in
\cite{Alvarez:CSF2004} that the power spectrum of
$|s(\mathbf{x},t)|$ can be used to estimate the value of $\omega$.
Moreover, the value of $\phi_0$ can be simultaneously obtained via
autocorrelation analysis \cite{WuHuZhang:CSF2004} or separately by
detecting zero-crossing points at intervals equal to
$\frac{\pi}{4\omega}$ \cite{Alvarez:CSF2004}. A direct extraction
of the binary plain-message signal from the transmitted signal has
also been discussed in \cite{Alvarez:CSF2004} for the chaotic
switching configuration.

To resist the proposed attacks, a modified modulating signal was
suggested in \cite{WuHuZhang:CSF2004}: $g(t)=A(\cos(\omega
t+\phi_0)+M)x_3(t)$, where $M>1$. It was claimed that such a
simple modification can eliminate the security defect of the
original scheme and can also improve the synchronization
performance. The default parameters of the modified scheme are
identical to the ones used in the original Bu-Wang scheme, except
$\phi_0=0.3$.

\section{Breaking the scheme based on the modified modulation method}
\label{section:cryptanalysis}

In this section, we show that the modified modulation method
proposed in \cite{WuHuZhang:CSF2004} can still be easily broken
via a new attack, which is designed based on parameter estimation
from the transmitted signal $s(\mathbf{x},t)$. It will be shown
that the proposed attack works for both chaotic masking and
chaotic switching configurations. The most significant feature of
the new attack is its independence of the zero-crossing points and
its robustness with respect to different values of $\omega$,
$\phi_0$ and $M$. This new attack is also able to break the
original Bu-Wang scheme. This paper further points out that the
modulation method itself is not very satisfactory from a
cryptographical point of view, even if the modulating signal
itself can be designed to be extremely secure against attacks.

\subsection{Breaking the values of $\omega$ and $\phi_0$}

\begin{figure}
\centering
\includegraphics[width=\textwidth]{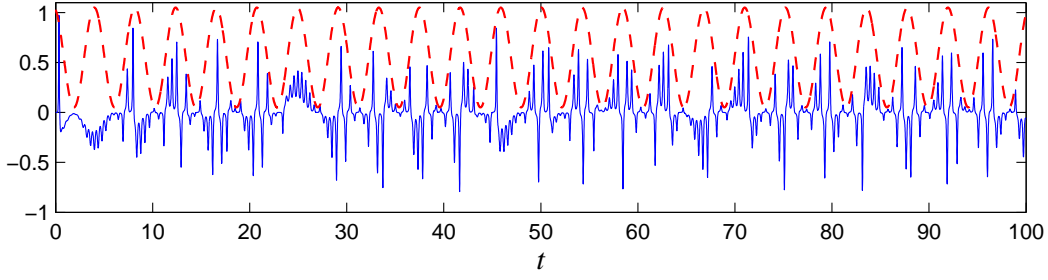}
a) $s(\mathbf{x},t)$ vs $g_0(t)=A(\cos(\omega t+\phi_0)+M)$, when
$i(t)=0$\\[\subfigskip]
\includegraphics[width=\textwidth]{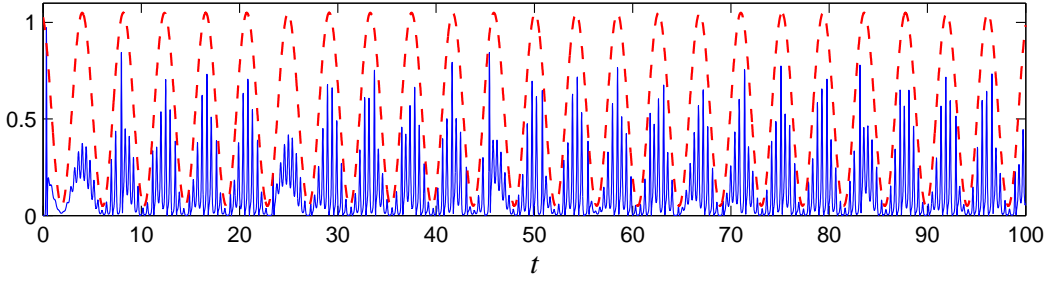}
b) $|s(\mathbf{x},t)|$ vs $g_0(t)=A(\cos(\omega
t+\phi_0)+M)$, when $i(t)=0$\\[\subfigskip]
\includegraphics[width=\textwidth]{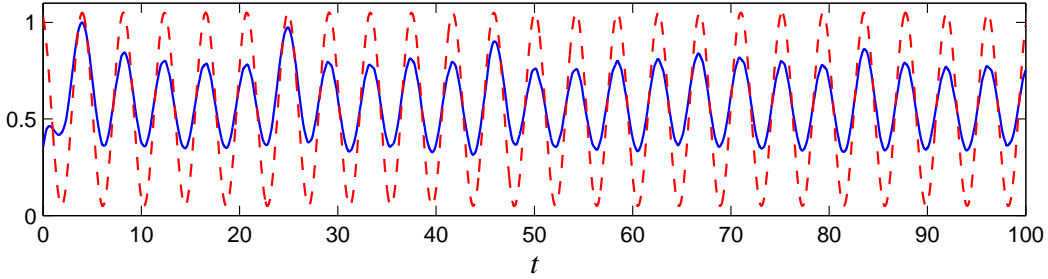}
c) $\bar{s}(\mathbf{x},t)$ vs $g_0(t)=A(\cos(\omega t+\phi_0)+M)$,
when $i(t)=0$\\[\subfigskip]
\includegraphics[width=\textwidth]{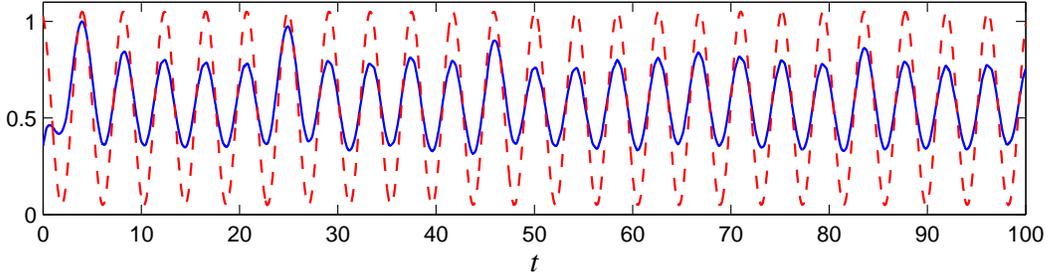}
d) $\bar{s}(\mathbf{x},t)$ vs $g_0(t)=A(\cos(\omega t+\phi_0)+M)$,
when $i(t)=0.1\sin(t)$%
\caption{Breaking the values of $\omega$ and $\phi_0$ in the
chaotic masking configuration, where and throughout the paper the
dashed line is $g_0(t)$.}\label{figure:attack_CMs}
\end{figure}

Assuming $g_0(t)=A(\cos(\omega t+\phi_0)+M)$, where $M>1$ implies
$g_0(t)>0$. Thus,
$|s(\mathbf{x},t)|=|g_0(t)x_3(t)x_1(t)|=g_0(t)|x_3(t)x_1(t)|$,
which means that the amplitude of $|x_3(t)x_1(t)|$ is periodically
modulated by $g_0(t)$ at a fixed frequency $\omega$. Consequently,
there exists an embedded periodicity in the amplitude of
$|s(\mathbf{x},t)|$, just like the embedded periodicity in the
zero-crossing points discussed in \cite{CheeXuBishop:CSF2004,
WuHuZhang:CSF2004, Alvarez:CSF2004}. See Figs.
\ref{figure:attack_CMs}a and b for a comparison of the waveforms
of $s(\mathbf{x},t)$, $|s(\mathbf{x},t)|$ and $g_0(t)$, in the
chaotic masking configuration with $i(t)=0$, where the waveforms
of $s(\mathbf{x},t)$ and $|s(\mathbf{x},t)|$ have been normalized
to make the display clearer. One can see that the periodicity
occurs as a clear envelop of the amplitude of $|s(\mathbf{x},t)|$.
After filtering $|s(\mathbf{x},t)|$ with an averaging filter, one
can dramatically refine the periodicity, as shown in Fig.
\ref{figure:attack_CMs}c, where
$\bar{s}(\mathbf{x},t)=|s(\mathbf{x},t)|\circledast f_a(t,4)$ has
also been normalized for a better display and the employed
averaging filter $f_a(t,h)$ is defined as follows:
\begin{equation}
f_a(t,h)=\begin{cases}
1-(|t|/h), & |t|\leq h,\\
0, & |t|>h.
\end{cases}
\end{equation}
Here, $\circledast$ denotes the convolution operation and the
value of $h$ should be chosen close to $\omega$ to get a smooth
filtered signal $\bar{s}(\mathbf{x},t)$. For this purpose, a rough
estimation of $\omega$ can be easily obtained from
$|s(\mathbf{x},t)|$ as shown in Fig. \ref{figure:attack_CMs}b.
Note that any other averaging filter may be used instead of the
above triangular one.

\begin{figure}
\centering
\includegraphics[width=\textwidth]{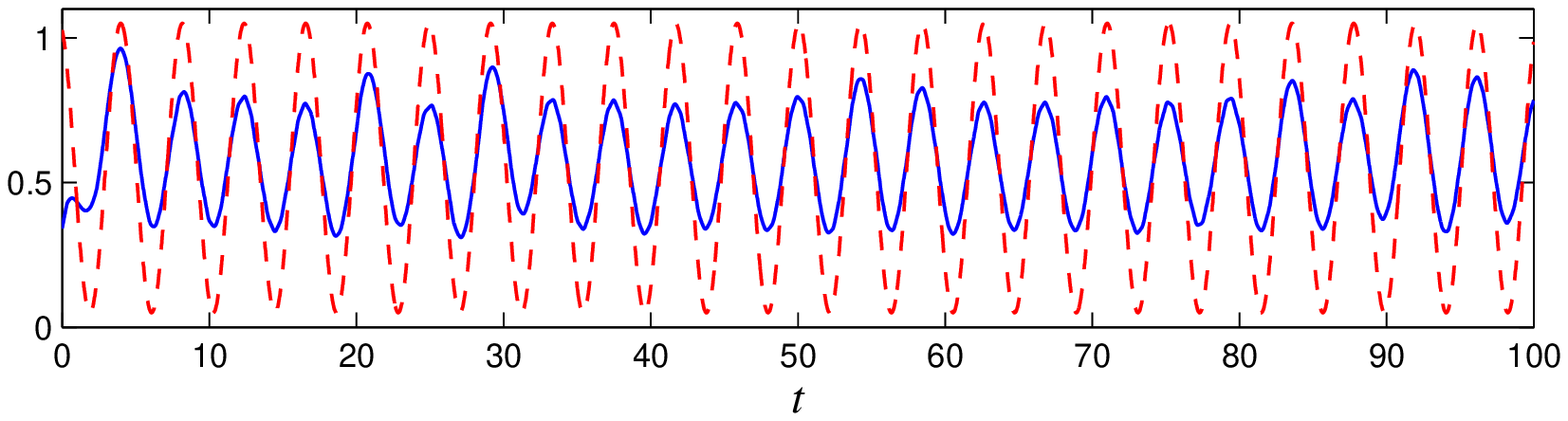}
\caption{Breaking the values of $\omega$ and $\phi_0$ in the
chaotic switching configuration: $\bar{s}(\mathbf{x},t)$ vs
$g_0(t)=A(\cos(\omega t+\phi_0)+M)$.}\label{figure:attack_CSKs}
\end{figure}

From Fig. \ref{figure:attack_CMs}c, it is clear that one can
easily get an accurate estimation of $\omega$ and $\phi_0$, thanks
to the global phase matching between $\bar{s}(\mathbf{x},t)$ and
$g_0(t)$. Of course, some other estimation algorithms, such as the
autocorrelation identification method proposed in
\cite{WuHuZhang:CSF2004}, can be employed to further improve the
accuracy of the estimation. When a plain-message signal is carried
in the transmitted signal $s(\mathbf{x},t)$, the above estimation
algorithm still works well as shown in Fig.
\ref{figure:attack_CMs}d, which is attributed to the sufficiently
small energy of the plain-message signal. Also, it is obvious that
the above algorithm works well for the chaotic switching
configuration (see Fig. \ref{figure:attack_CSKs} for a breaking
result).

\begin{figure}
\centering
\includegraphics[width=\textwidth]{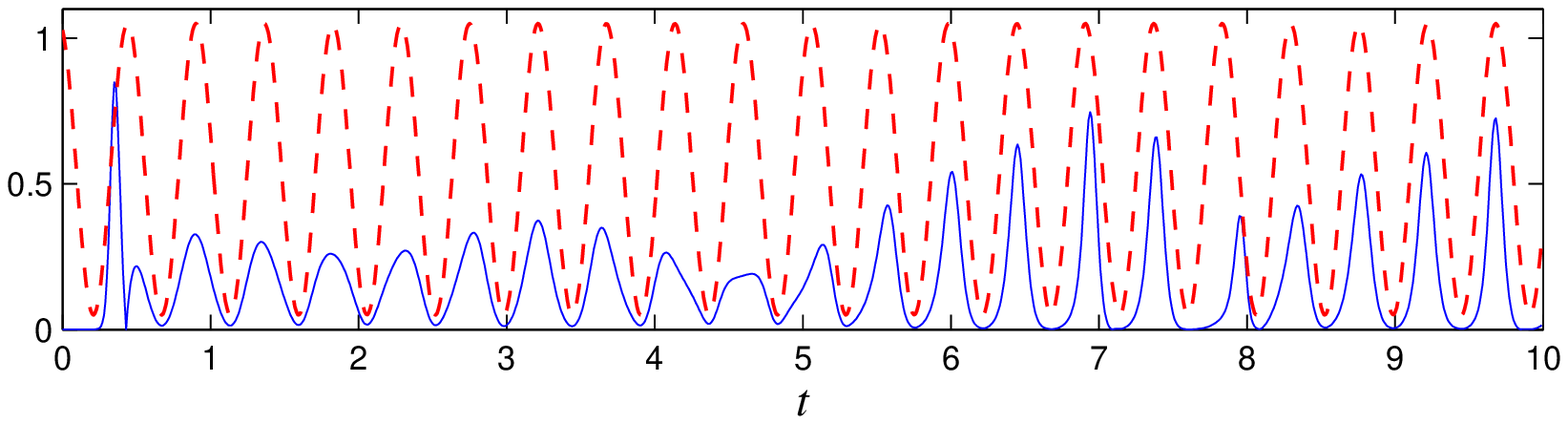}
a) $|s(\mathbf{x},t)|$ vs $g_0(t)=A(\cos(\omega
t+\phi_0)+M)$\\[\subfigskip]
\includegraphics[width=\textwidth]{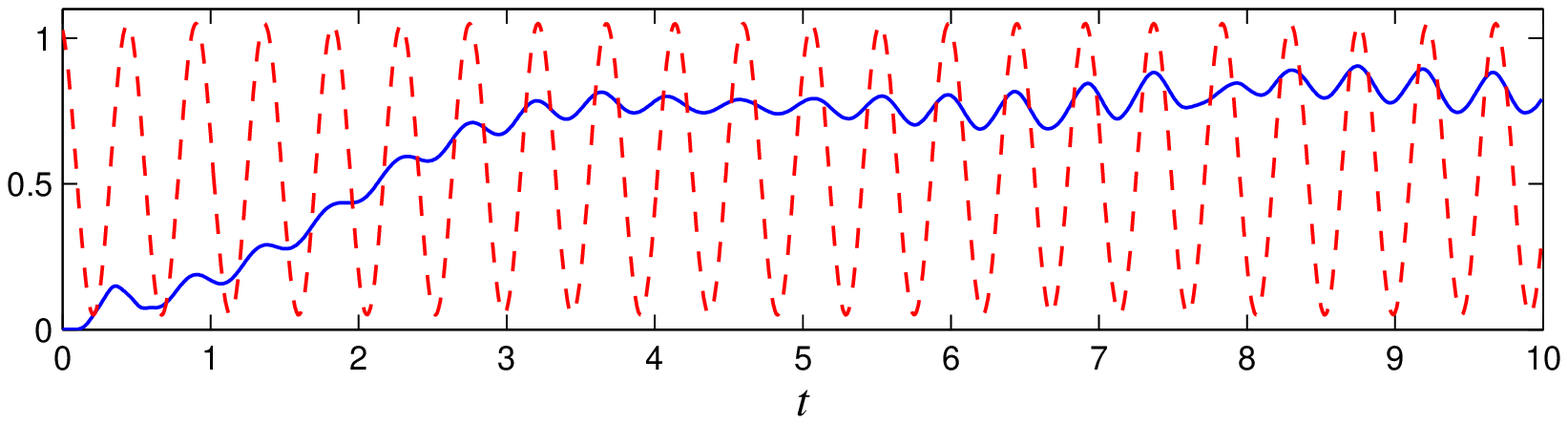}
b) $\bar{s}(\mathbf{x},t)$ vs $g_0(t)=A(\cos(\omega t+\phi_0)+M)$
\caption{Breaking the values of $\omega$ and $\phi_0$ in the
chaotic masking configuration.}\label{figure:attack_CMs2}
\end{figure}

From the above discussion, it seems that the performance of the
proposed estimation will gradually decrease as $\omega$ increases,
since the averaging filter $f_a(t,h)$ may produce unsatisfactory
results. This looks like a defect of the proposed attack. However,
our experiments show that when $\omega$ is relatively large, it
becomes possible to directly estimate the values of $\omega$ and
$\phi_0$ from $|s(\mathbf{x},t)|$ only (even without using average
filtering). When $\omega=13.6$\footnote{This value is
intentionally set to be identical to the inherent frequency $2\pi
f_z$ of the Lorenz system, so as to emphasize on the robustness of
the proposed attack. Some explanations of the inherent frequency
of the Lorenz system are given in the following paragraphs.}, for
example, the waveforms of $|s(\mathbf{x},t)|$,
$\bar{s}(\mathbf{x},t)$ and $g_0(t)$ are shown in Fig.
\ref{figure:attack_CMs2}, where the parameter of the averaging
filter is chosen as $h=0.2$. Note that the positions of local
minima of $|s(\mathbf{x},t)|$ coincide very well with those of
$g_0(t)$, although $\bar{s}(\mathbf{x},t)$ becomes less clear.
This demonstrates that the proposed attack method is rather robust
to different values of $\omega$ and $\phi_0$.

\begin{figure}
\centering
\includegraphics[width=\textwidth]{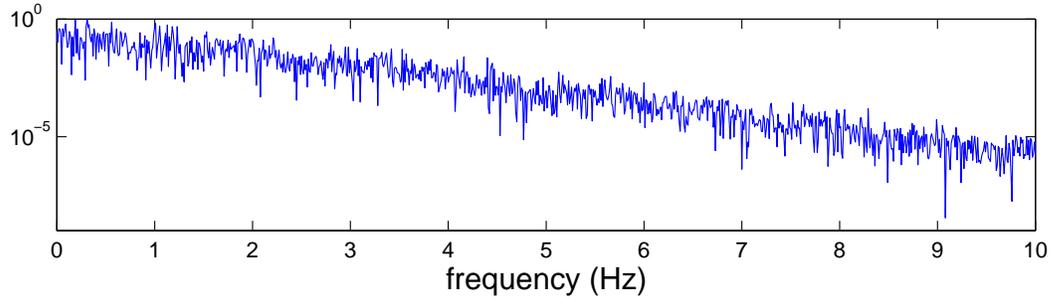}
a) the relative power spectrum of $x_1(t)$\\
\includegraphics[width=\textwidth]{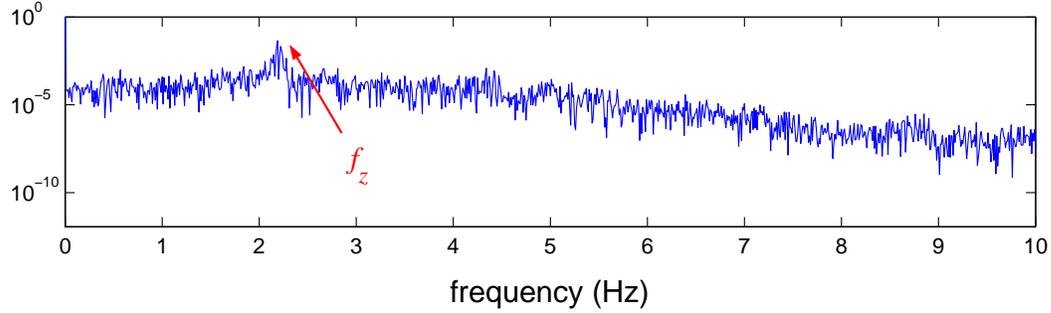}
b) the relative power spectrum of $|x_1(t)|$\\
\includegraphics[width=\textwidth]{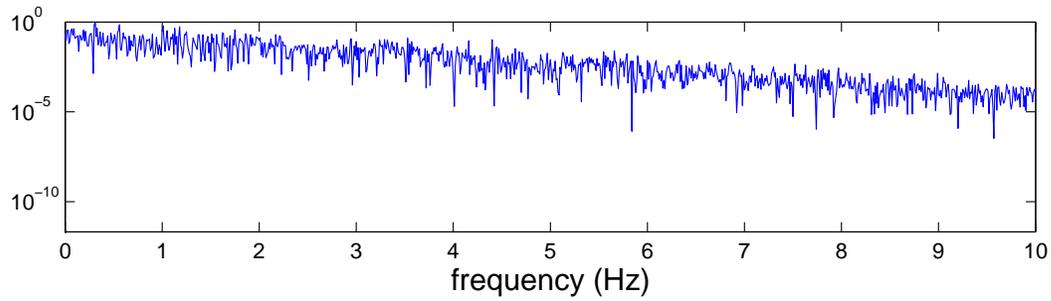}
c) the relative power spectrum of $s(\mathbf{x},t)$\\
\includegraphics[width=\textwidth]{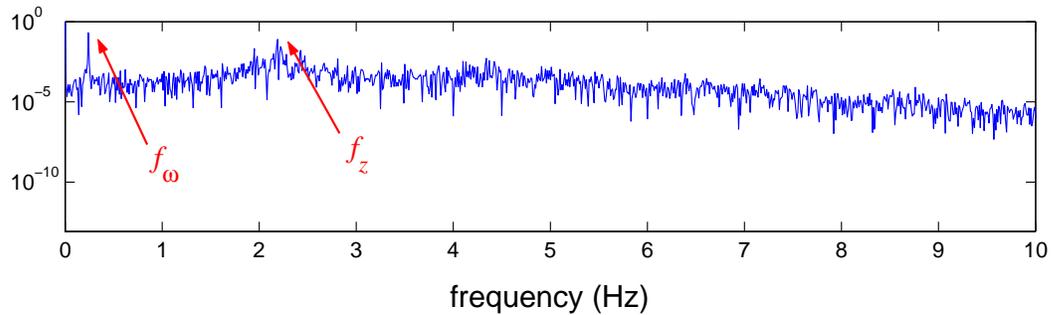}
d) the relative power spectrum of $|s(\mathbf{x},t)|$%
\caption{The relative power spectra of $x_1(t)$, $|x_1(t)|$,
$s(\mathbf{x},t)$ and
$|s(\mathbf{x},t)|$.}\label{figure:InherentFrequency}
\end{figure}

It is worth mentioning that the value of $\omega$ can also be
derived via spectral analysis: simply, calculating the power
spectrum of $\bar{s}(\mathbf{x},t)$ and extracting the spectrum
peak at the frequency $\omega$. In fact, such a direct spectral
analysis for estimating $\omega$ works even without using average
filtering, as shown in \cite{Alvarez:CSF2004}. This is because the
following property of Lorenz-like chaotic systems
\cite{Alvarez-Li2004b, Alvarez-Li2004c}: although $x_1(t)$ of the
Lorenz system has a wideband spectrum as shown in Fig.
\ref{figure:InherentFrequency}a, the absolute-valued signal
$|x_1(t)|$ has a prominent spectrum peak at an inherent frequency
$f_z$, which is uniquely determined by the system parameters
$\sigma$, $b$ and $r$ and is independent of the initial
conditions. See Fig. \ref{figure:InherentFrequency}b for the power
spectrum of $|x_1(t)|$. This is also true for $x_2(t)$. For
$x_3(t)$, the operation is even simpler: $x_3(t)$ itself has a
prominent spectrum peak at the inherent frequency $f_z$. This
implies that $|s(\mathbf{x},t)|=g_0(t)|x_3(t)x_1(t)|$ will have at
least two prominent spectrum peaks, at $f_z$ and
$f_\omega=\dfrac{\omega}{2\pi}$ (Hz), respectively. See Figs.
\ref{figure:InherentFrequency}c and d for a comparison of the
power spectra of $s(\mathbf{x},t)$ and $|s(\mathbf{x},t)|$. It can
be seen that the first prominent peak is at the frequency
$f_\omega$ and the second at $f_z$. Apparently, $\omega$ can also
be obtained from $f_\omega$ as $\omega=2\pi f_\omega$. Following
\cite{Alvarez-Li2004b, Alvarez-Li2004c}, such an inherent
frequency also exists in Chua and R\"{o}ssler systems, so the same
spectral analysis is also available for attacking these two
chaotic systems widely-used in chaos-based secure communications.
At present, it is not clear whether there exists any chaotic
system that has no distinguishable inherent frequency.

Compared with the spatial attack proposed above, the existing
spectral attack has some defects: 1) it cannot work well when
$\omega\approx f_z$; 2) the value of $\phi_0$ cannot be
simultaneously estimated with $\omega$ in the frequency domain; 3)
if there does not exist a strong DC component in $|x_3(t)x_1(t)|$
(e.g., if some other chaotic systems are also used), $f_\omega$
will not occur as a prominent peak in the power
spectrum\footnote{Although $f_z\pm f_\omega$ always occur as two
minor peaks, they are generally less distinguishable than the
individual major peaks at $f_\omega$ and $f_z$, respectively.}; 4)
when other chaotic systems are used instead of the Lorenz system,
the inherent frequency may not exist (though unlikely) or the
frequency $f_\omega$ might not be easily identified from the power
spectra.

\begin{figure}
\centering
\includegraphics[width=\textwidth]{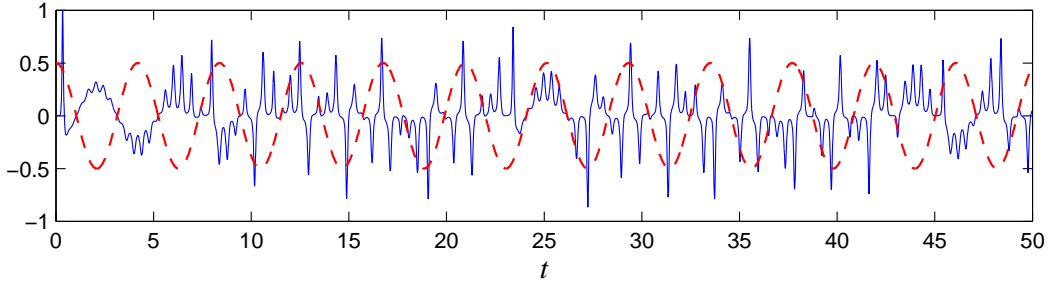}
a) $s(\mathbf{x},t)$ vs $g_0(t)=A\cos(\omega t+\phi_0)$
\\[\subfigskip]
\includegraphics[width=\textwidth]{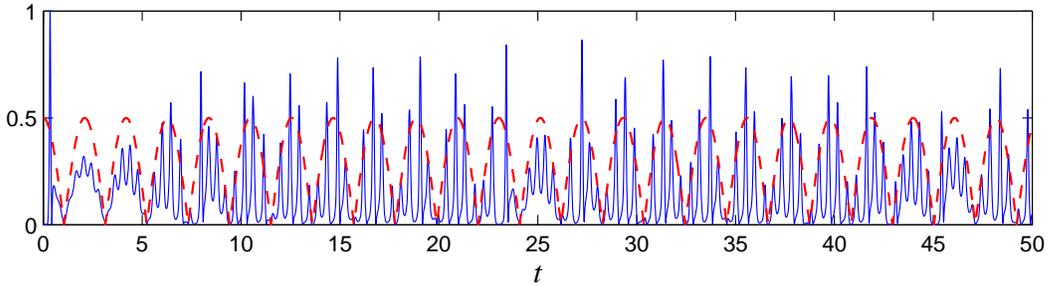}
b) $|s(\mathbf{x},t)|$ vs $|g_0(t)|=|A\cos(\omega
t+\phi_0)|$\\[\subfigskip]
\includegraphics[width=\textwidth]{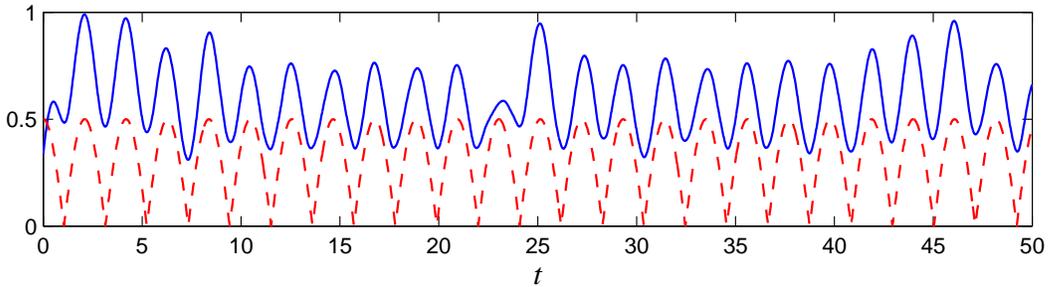}
c) $\bar{s}(\mathbf{x},t)$ vs $|g_0(t)|=|A\cos(\omega
t+\phi_0)|$%
\caption{Breaking the values of $\omega$ and $\phi_0$ in the
original Bu-Wang scheme, where the dashed line denotes either
$g_0(t)$ or $|g_0(t)|$.}\label{figure:attackBW_CMs}
\end{figure}

Finally, let us see how the proposed attack works on the original
Bu-Wang scheme, where the condition is somewhat different since
$g_0(t)=A\cos(\omega t+\phi_0)$ is not always positive. In that
case, one has
\begin{eqnarray*}
|s(\mathbf{x},t)| & = & A|\cos(\omega
t+\phi_0)|\cdot|x_3(t)x_1(t)|\\
& = & \frac{A}{\sqrt{2}}\cdot\sqrt{1+\cos(2\omega
t+2\phi_0)}\cdot|x_3(t)x_1(t)|.
\end{eqnarray*}
Thus, with the above-described attack we will get $2\omega$ and
$2\phi_0$, which can be easily halved to obtain the values of
$\omega$ and $\phi_0$. In Fig. \ref{figure:attackBW_CMs}, the
result of breaking the chaotic masking configuration with $i(t)=0$
of the original Bu-Wang scheme is shown, where a different
averaging filter, $f_a'(t,1)$, is used to make
$\bar{s}(\mathbf{x},t)$ smoother:
\begin{equation}
f_a'(t,h)=\begin{cases}
1+\cos\left(\frac{2\pi t}{2h}\right), & |t|\leq h\\
0, & |t|>h.
\end{cases}
\end{equation}

From the above analysis, one can see that the attack proposed in
this subsection can work for not only the original Bu-Wang scheme
but also some other scheme if the modulating signal $g_0(t)$ is
periodic (which need not be a simple sinusoidal or cosine signal).
Therefore, an aperiodic or even chaotic modulating signal $g_0(t)$
has to be used to enhance the security. However, in this case the
synchronization of the modulating signal itself becomes a new
problem, which will be further discussed in
Sec.~\ref{section:performance}.

\subsection{Breaking the value of $M$}

\begin{figure}
\centering
\begin{minipage}{\figwidth}
\centering
\includegraphics[width=\textwidth]{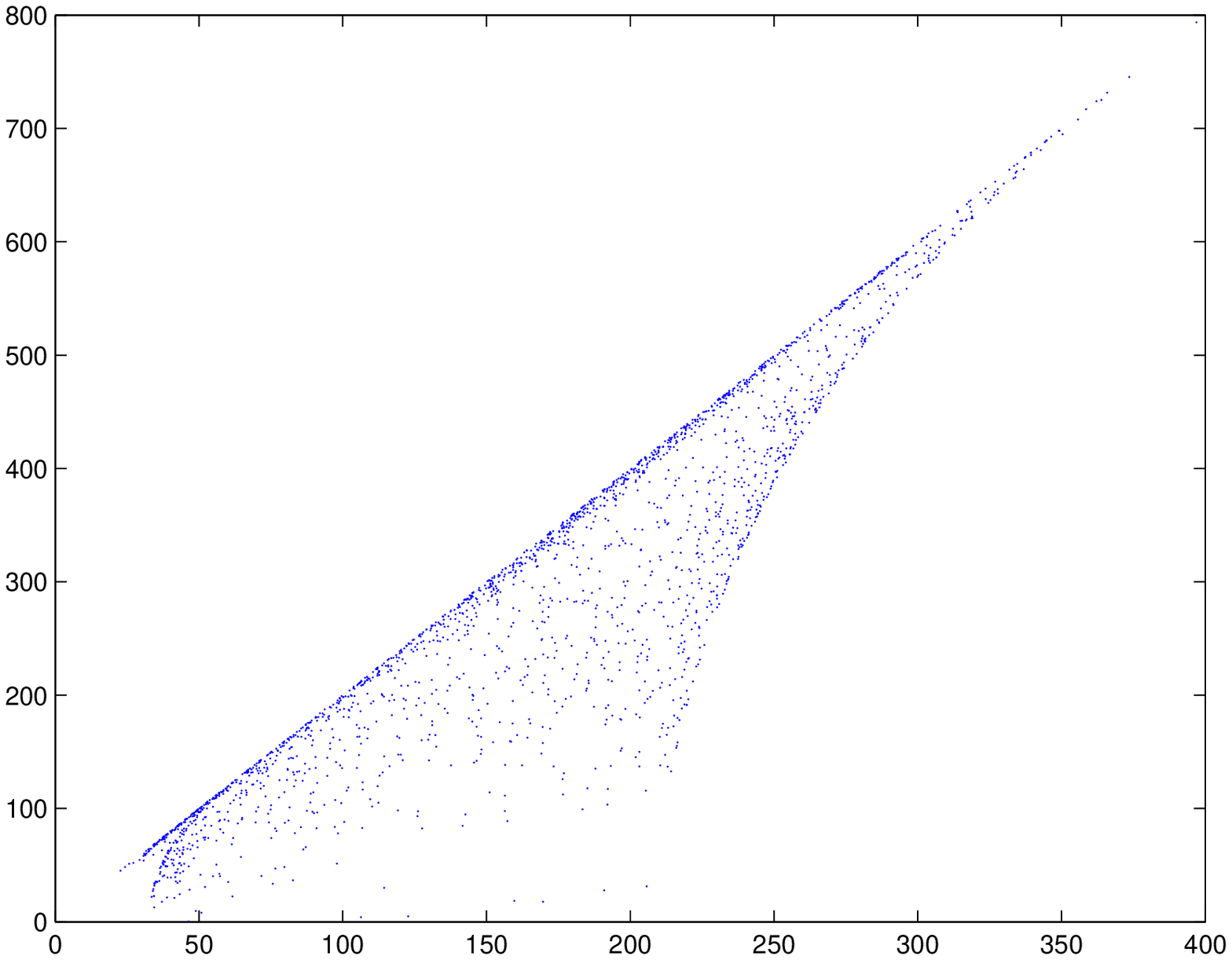}
a) $M'=M+0.9$
\end{minipage}
\begin{minipage}{\figwidth}
\centering
\includegraphics[width=\textwidth]{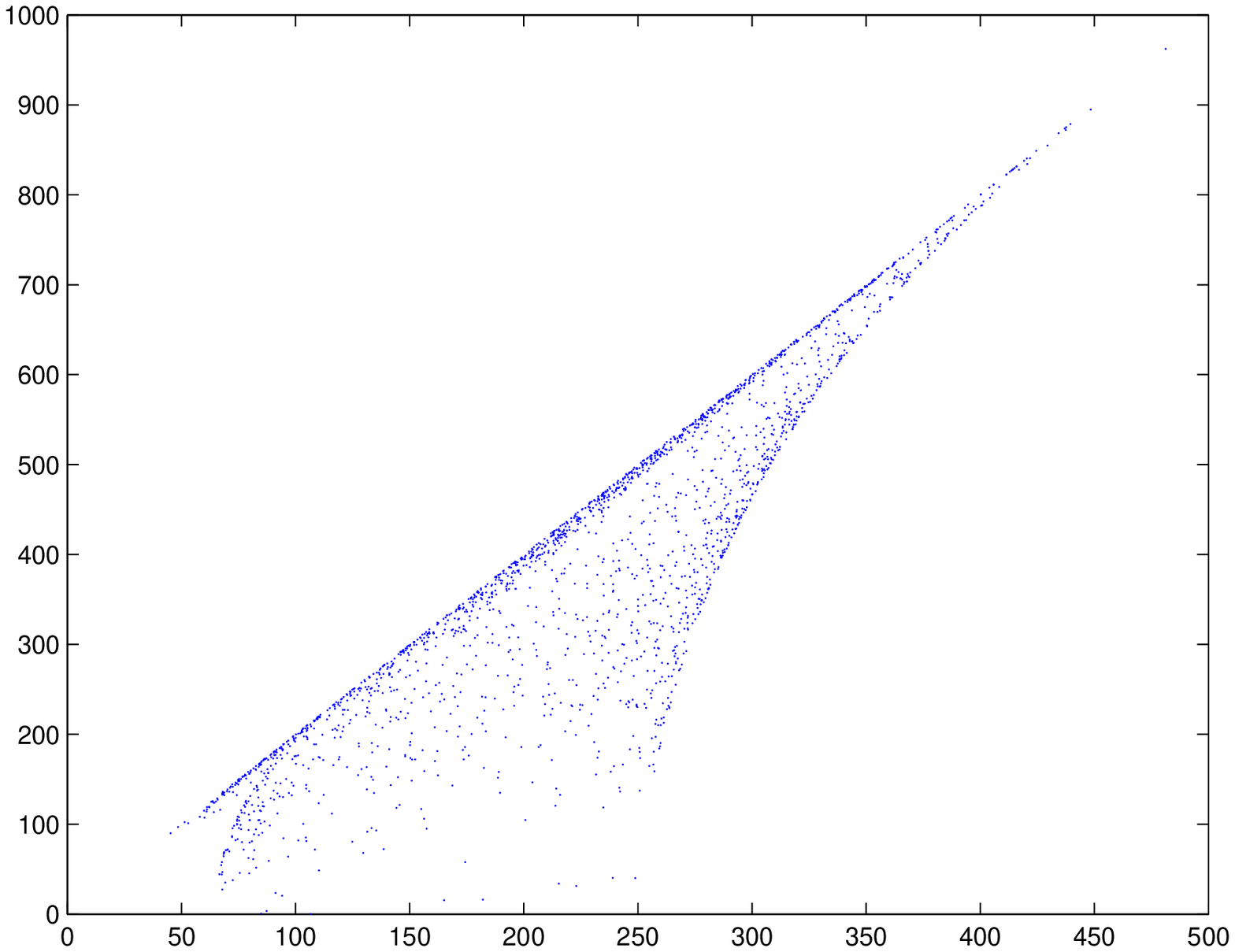}
b) $M'=M+0.4$
\end{minipage}
\begin{minipage}{\figwidth}
\centering
\includegraphics[width=\textwidth]{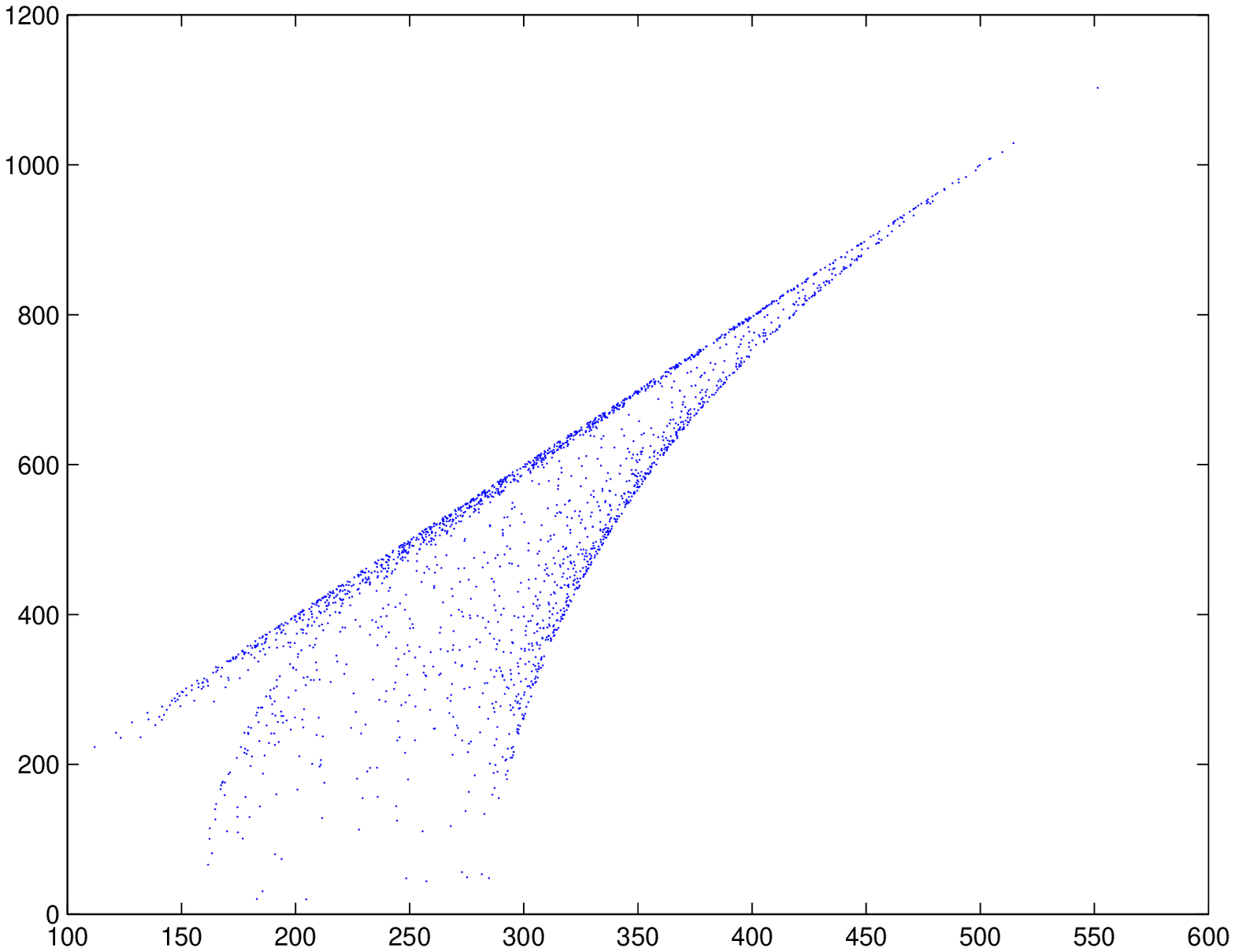}
c) $M'=M+0.1$
\end{minipage}\\[\subfigskip]
\begin{minipage}{\figwidth}
\centering
\includegraphics[width=\textwidth]{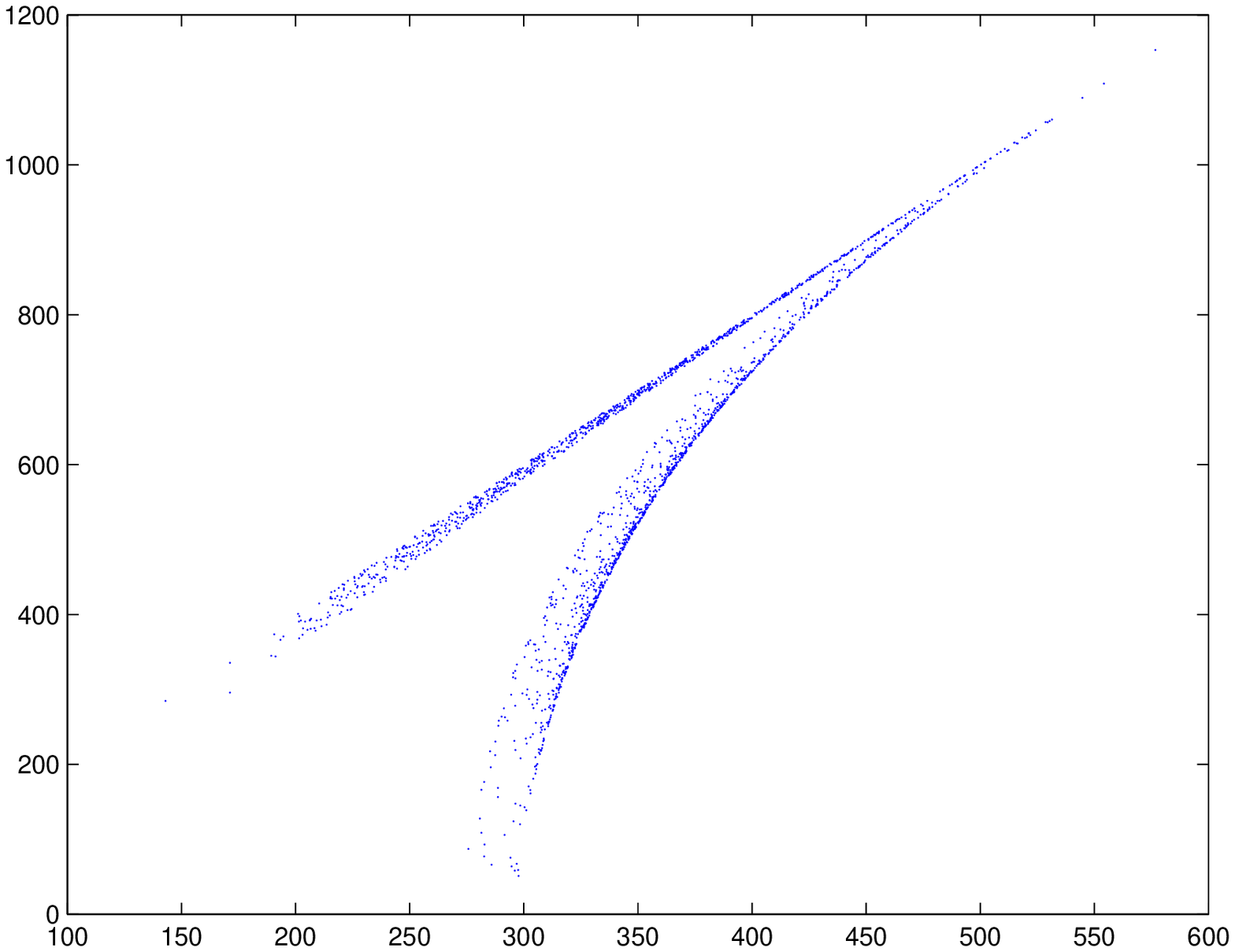}
d) $M'=M+0.01$
\end{minipage}
\begin{minipage}{\figwidth}
\centering
\includegraphics[width=\textwidth]{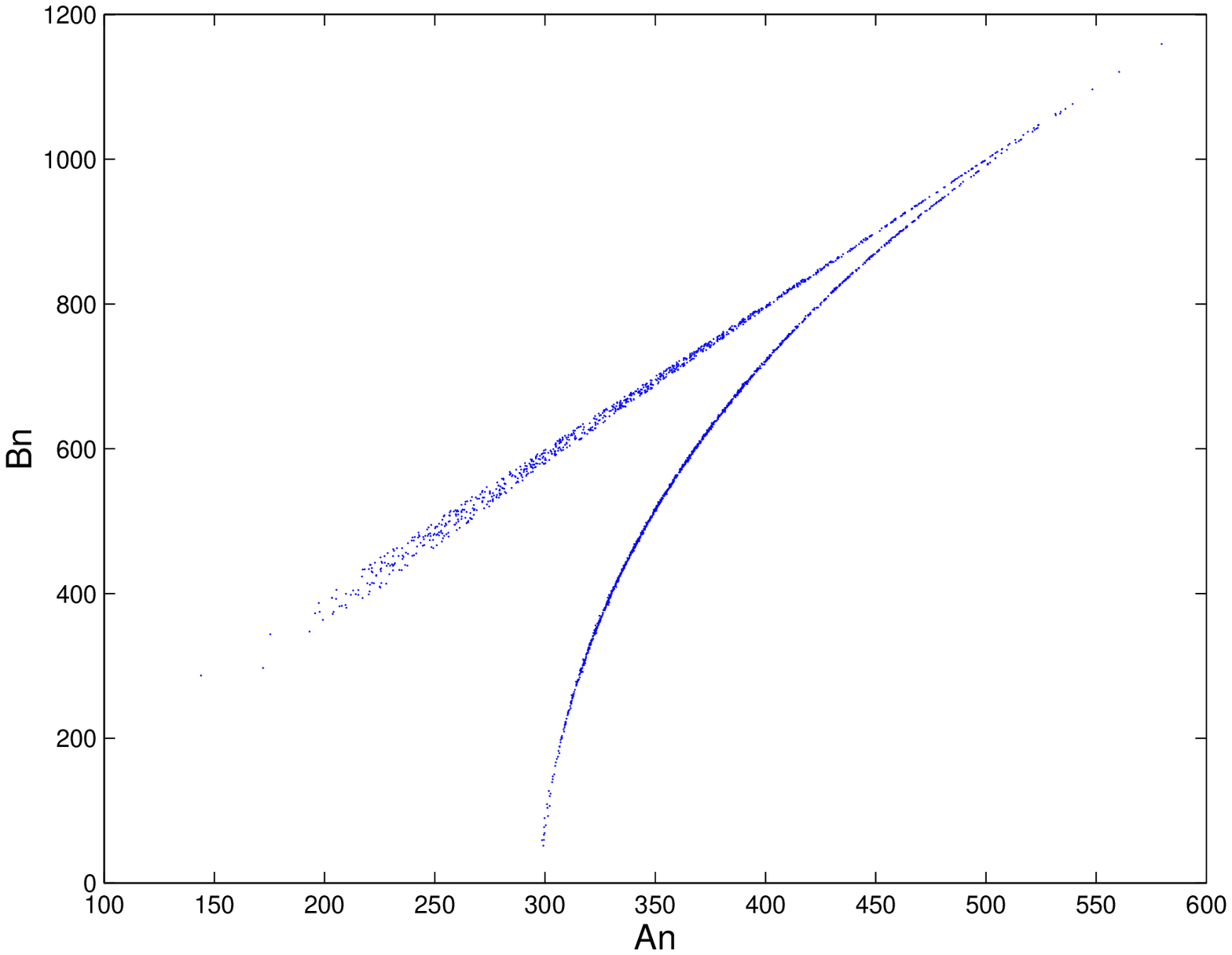}
e) $M'=M$
\end{minipage}
\begin{minipage}{\figwidth}
\centering
\includegraphics[width=\textwidth]{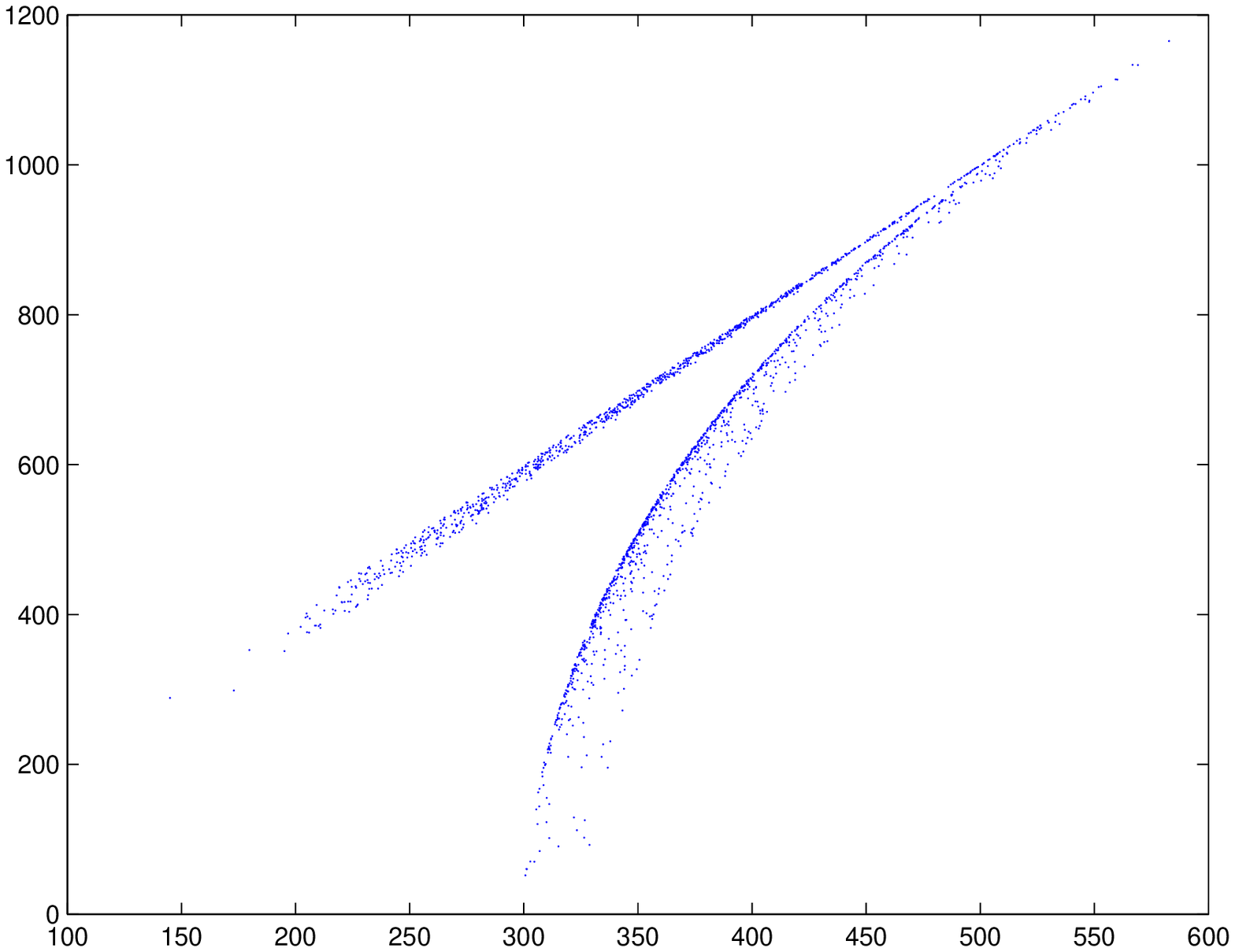}
f) $M'=M-0.01$
\end{minipage}\\[\subfigskip]
\begin{minipage}{\figwidth}
\centering
\includegraphics[width=\textwidth]{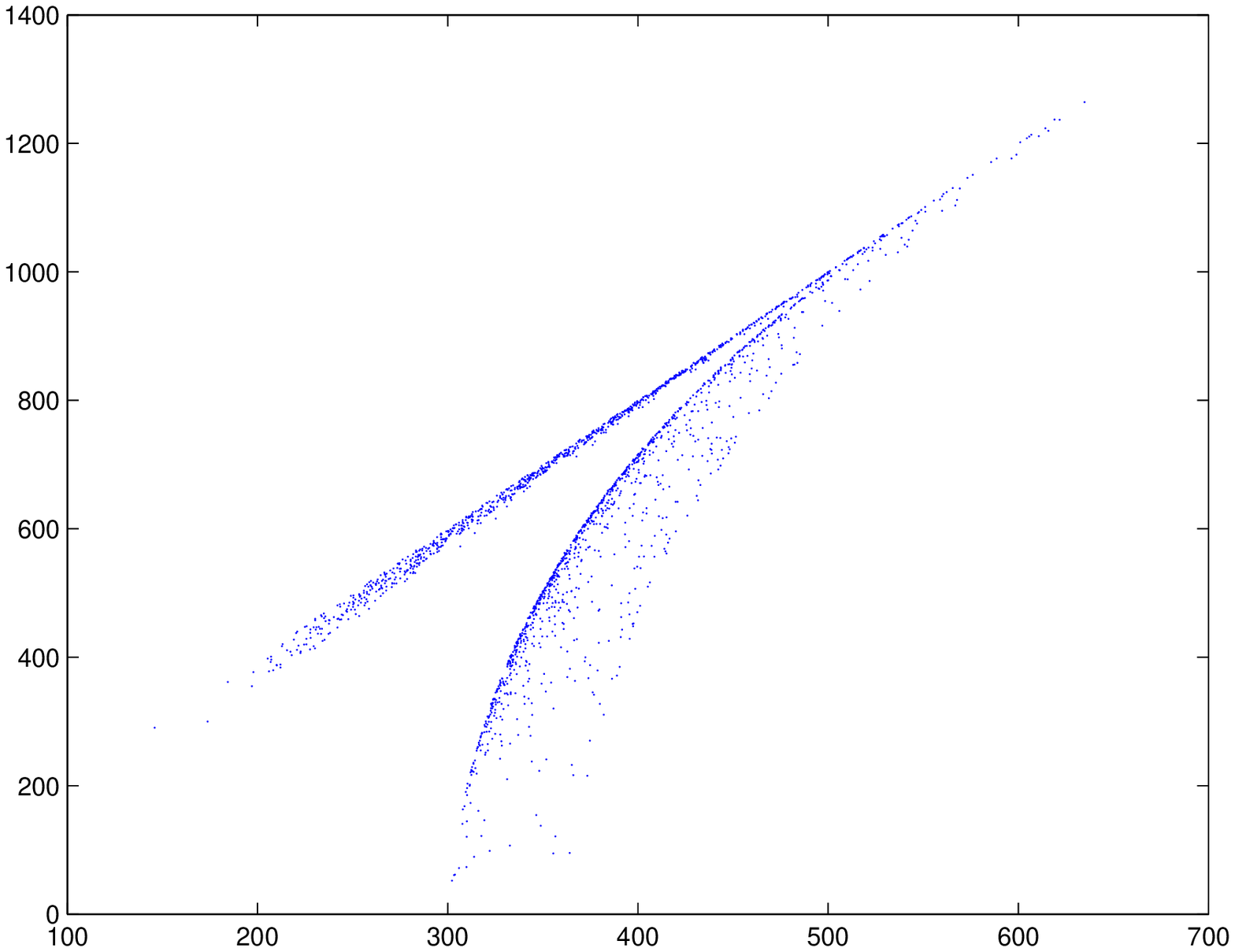}
g) $M'=M-0.02$
\end{minipage}
\begin{minipage}{\figwidth}
\centering
\includegraphics[width=\textwidth]{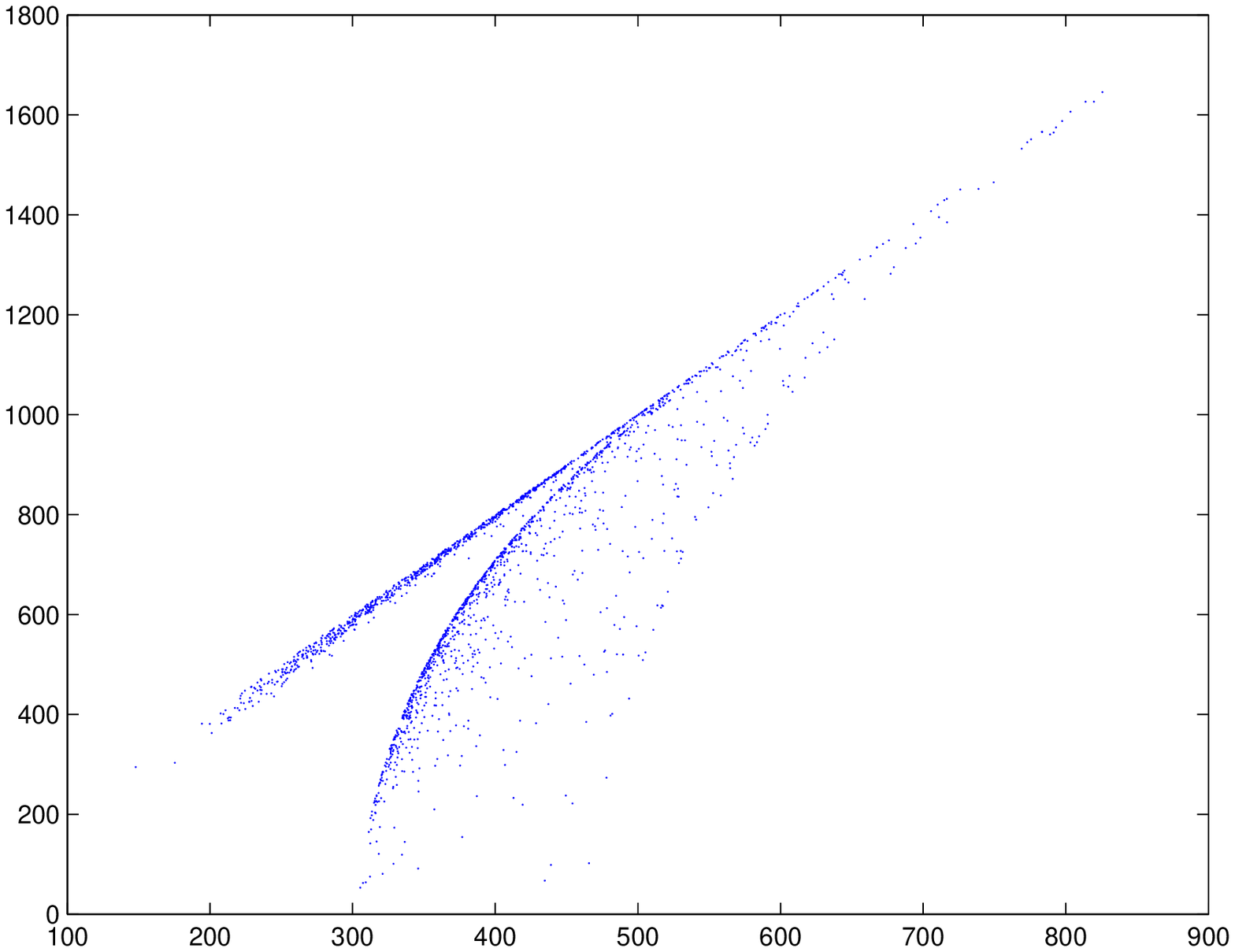}
h) $M'=M-0.04$
\end{minipage}
\begin{minipage}{\figwidth}
\centering
\includegraphics[width=\textwidth]{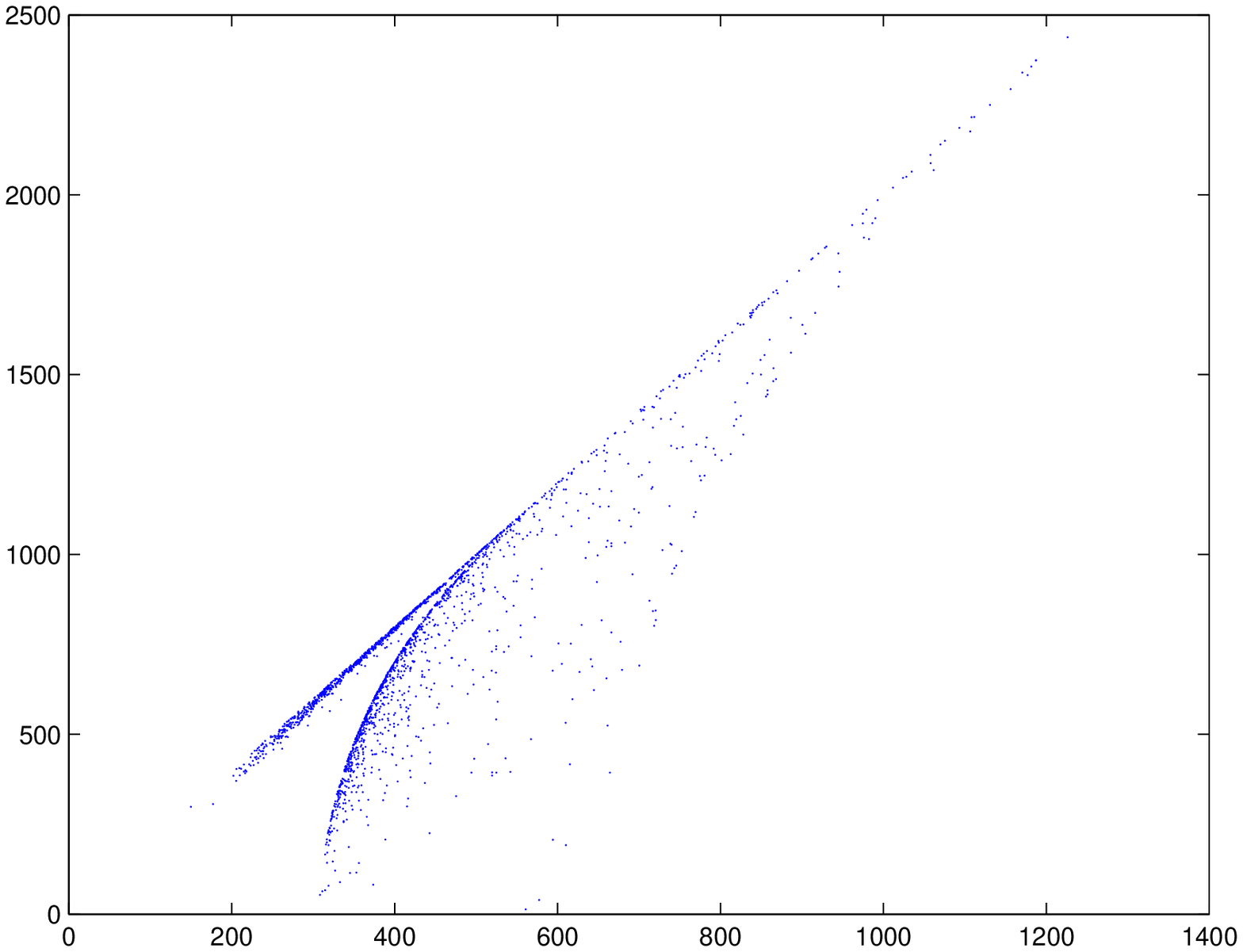}
i) $M'=M-0.06$
\end{minipage}
\caption{Return maps reconstructed from $|\hat{s}(\mathbf{x},t)|$
with different values of $M'$.}\label{figure:ReturnMap1}
\end{figure}

Once the values of $\omega$ and $\phi_0$ are known, one can
further break the value of $M$ via an iterative process, and then
get rid of the blurring effect on the return map by removing
$g_0(t)$ from $s(\mathbf{x},t)$. To do so, define an auxiliary
signal
\begin{equation}
\hat{s}(\mathbf{x},t)=\frac{s(\mathbf{x},t)}{\cos(\omega
t+\phi_0)+M'}=\frac{\cos(\omega t+\phi_0)+M}{\cos(\omega
t+\phi_0)+M'}\cdot Ax_3(t)x_1(t),
\end{equation}
where $M'$ is an estimation of $M$. Apparently, it is expected
that the closer the value of $M'$ is to $M$, the closer the signal
$\hat{s}(\mathbf{x},t)$ is to $\hat{x}_{13}(t)=Ax_3(t)x_1(t)$. As
a natural result, if $M'$ changes from a value larger than $M$ to
a value smaller than $M$, i.e., $(M'>M)\to(M'<M)$, the behavior of
$\hat{s}(\mathbf{x},t)$ will gradually go closer to
$\hat{x}_{13}(t)$ and then turn gradually far away from
$\hat{x}_{13}(t)$ once $M'$ crosses the point $M'=M$. Therefore,
there exists a global minimum at the point $M'=M$. The existence
of such a global minimum can be easily checked by observing the
reconstructed return map from $\hat{s}(\mathbf{x},t)$ or
$|\hat{s}(\mathbf{x},t)|$. In Fig. \ref{figure:ReturnMap1}, the
return maps constructed from $|\hat{s}(\mathbf{x},t)|$ with
respect to different values of $M'$ are shown. The reason why
$|\hat{s}(\mathbf{x},t)|$ is used is that this reconstructed
return map has a simpler structure than the map reconstructed from
$\hat{s}(\mathbf{x},t)$. The following features can be found from
the maps:
\begin{enumerate}
\item The closer the value of $M'$ is to $M$, the thinner the
branch width and the closer the return map reconstructed from
$|\hat{s}(\mathbf{x},t)|$ is to the return map from
$|\hat{x}_{13}(t)|$.

\item The shape of the return map is related to the relation
between $M'$ and $M$: the two edges of the right-bottom branch are
both clear when $M'>M$, while only the left edge is clear when
$M'<M$. When $M'$ is closer to $M$, the above characteristic is a
little different: the right edge is (not so much) clearer than the
left one when $M'>M$, while the left edge is \textit{much} clearer
than the right one when $M'<M$.
\end{enumerate}

The first feature above means that one can approximately know the
distance between $M'$ and $M$ by measuring the width of the
right-bottom branch of the return map, and the second feature
means that one can exactly know whether $M'>M$ or $M'<M$. Note
that the branch width and the clearness of the two edges of the
right-bottom branch can be quantitatively defined by different
methods. For example, some algorithms can be used to extract the
skeleton and the two edges of the branch, and then determine the
two measures. In fact, in real attacks, an attacker can directly
get a rough estimation of the measures by naked eye, and then
\textit{manually} carry out the following iterative algorithm for
several times to find a sufficiently accurate estimation of $M$.

Given a lower bound $M_-<M$ and an upper bound $M_+>M$, the
iterative algorithm is as follows:
\begin{itemize}
\item \textit{Step 1 -- Initialization}: Set $i=1$,
$M_-^{(i)}=M_-$, $M_+^{(i)}=M_+$, and determine the number of
iterative rounds, $m$ (which is used to achieve a desired
precision, as further explained later).

\item \textit{Step 2}: Among the following $n+1$ values,
\begin{equation}
\left\{M'_{i,j}=M_-^{(i)}+j\cdot\frac{\left(M_+^{(i)}-M_-^{(i)}\right)}n\right\}_{j=0}^n,
\end{equation}
find two consecutive ones satisfying $M'_{i,j-1}<M$ and
$M'_{i,j}>M$.

\item \textit{Step 3}: If $i\leq m$, then increase $i$ by 1, set
$M_-^{(i)}=M'_{i,j-1}$, $M_+^{(i)}=M'_{i,j}$, and goto Step 2;
otherwise, stop the algorithm and output an estimation
$\widetilde{M}=\frac{M'_{i,j-1}+M'_{i,j}}{2}$.
\end{itemize}

Apparently, the above iterative algorithm is rather similar to the
numerical algorithms for solving algebraic equations for roots
\cite{Schatzman:NumerticalAnalysis}. It can be easily deduced that
the estimation accuracy after $m$ rounds is
$\left|\widetilde{M}-M\right|\leq\frac{M_+-M_-}{2n^m}$. To achieve
a predefined accuracy $\epsilon>0$, it is required that
$\frac{M_+-M_-}{2n^m}\leq\epsilon$, and then one gets
$m\geq\frac{\log_2(M_+-M_-)-\log_2(2\epsilon)}{\log_2n}$. Taking
$m=\left\lceil\frac{\log_2(M_+-M_-)-\log_2(2\epsilon)}{\log_2n}\right\rceil$,
the complexity of the above iterative algorithm is only
$O(nm)=O\left(n\cdot\frac{\log_2(M_+-M_-)-\log_2(2\epsilon)}{\log_2n}\right)$,
which is sufficiently small for an attacker to carry out the
algorithm \textit{manually}. When $n=3$, the computing complexity
is minimized, since the integer function $f(n)=\frac{n}{\log_2n}$
($n\geq 2$) reaches the global minimum at $n=3$.

\begin{figure}
\centering\setlength\figwidth{0.45\textwidth}
\begin{minipage}{\figwidth}
\centering
\includegraphics[width=\textwidth]{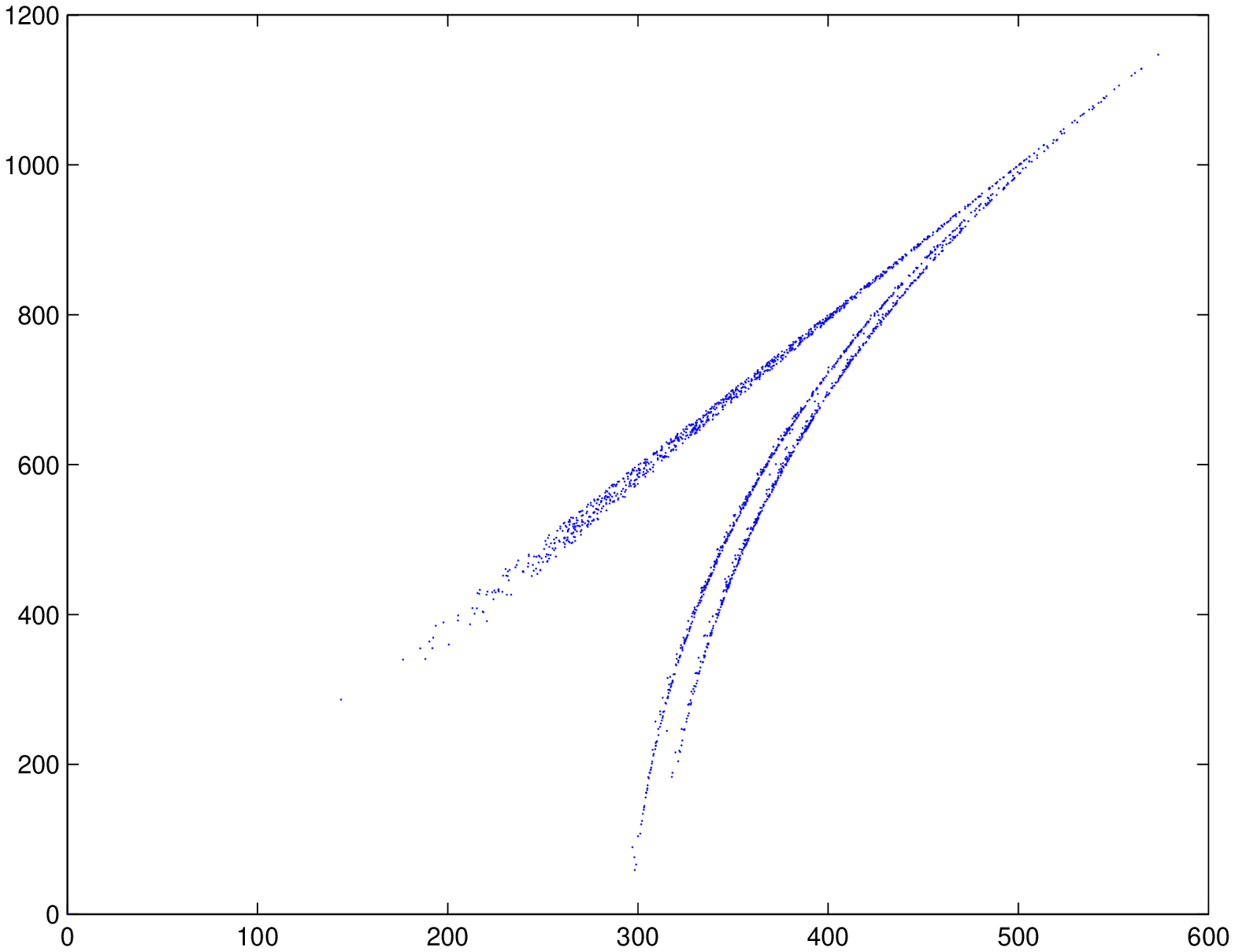}
a) The return map reconstructed from $|\hat{s}(\mathbf{x},t)|$
\end{minipage}
\begin{minipage}{\figwidth}
\centering
\includegraphics[width=\textwidth]{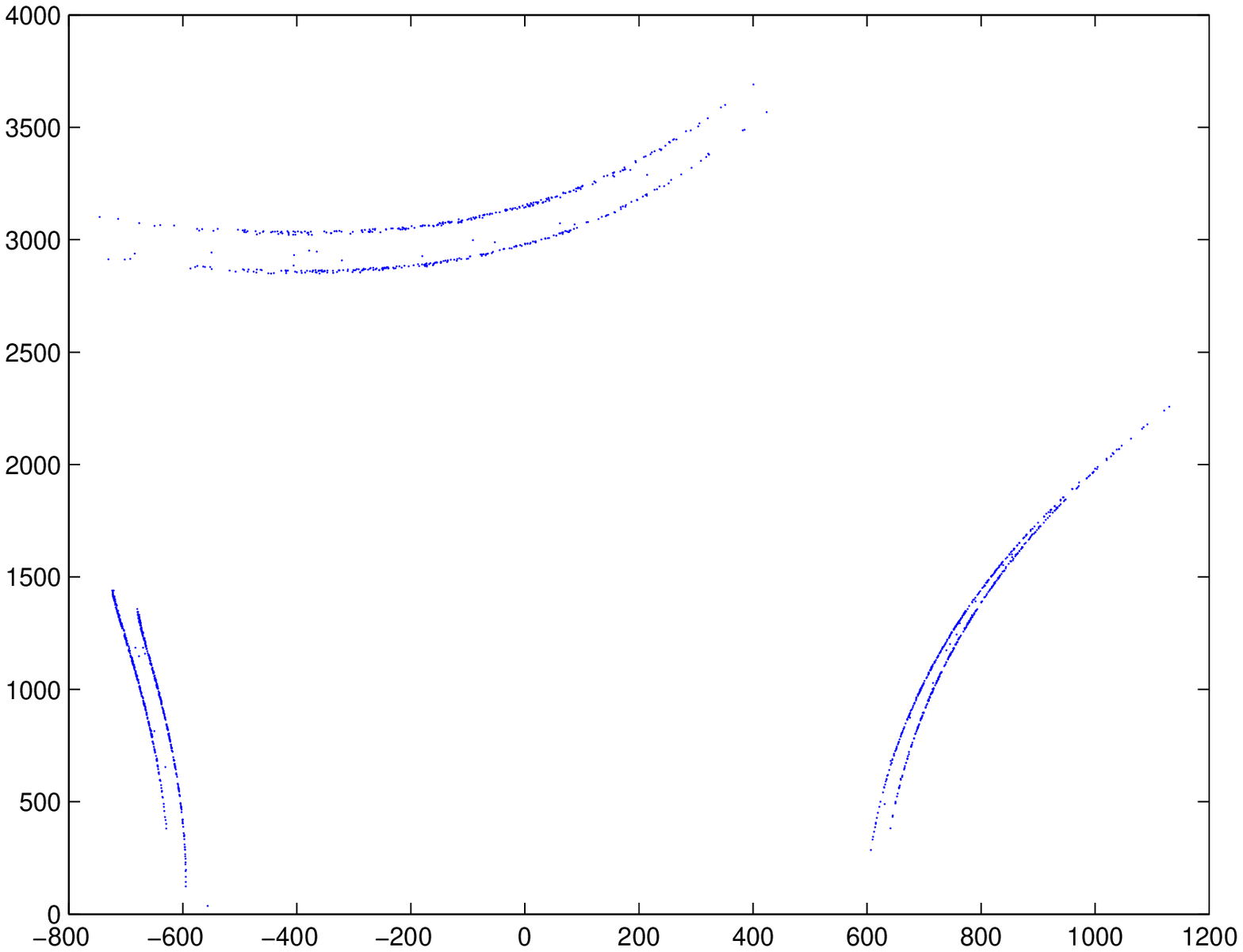}
b) The return map reconstructed from $\hat{s}(\mathbf{x},t)$
\end{minipage}
\caption{The return maps reconstructed from
$|\hat{s}(\mathbf{x},t)|$ and $\hat{s}(\mathbf{x},t)$ when
$M'=M+0.01$ in the chaotic switching
configuration.}\label{figure:ReturnMap2}
\end{figure}

Although in theory the above algorithm can get an arbitrarily
accurate estimation by increasing the round number $m$, the
distinguishability of the return map reconstructed from a finite
set of data will be insufficient when $M'$ is too close to $M$.
Fortunately, in this case, the return map will be sufficiently
clear for an intruder to carry out the return-map-based attacks.
Figure \ref{figure:ReturnMap2} shows the return map reconstructed
from $|\hat{s}(\mathbf{x},t)|$ and $\hat{s}(\mathbf{x},t)$ when
$M'=M+0.01$ in the chaotic switching configuration. It can be seen
that such an accuracy is enough to break the 0/1-bits in the
plain-message signal.

\section{Discussions on the performance of the modulation-based methods}
\label{section:discussions}

\subsection{A cryptographical perspective}

From a cryptographical point of view, a modulating signal is not
an effective tool for enhancing the security of chaos-based secure
communications, due to the dependence of the secret parameters of
the chaotic systems on the secret parameters of the modulating
signal. In fact, to break the whole chaotic cryptosystem via a
brute-force attack, what an attacker should do is to exhaustively
guess the secret parameters of the modulating signal, not all
secret parameters (system parameters and modulating parameters).
This means that the key space of the whole system is reduced to be
$\max(K_m,K_c)$, i.e., to be $K_m$ when $K_m>K_c$, where $K_m$ is
the key space of the modulating signal and $K_c$ is the key space
of the chaotic cryptosystem without modulation. However, for a
good encryption algorithm, the key space should be always
$K_m\times K_c$ \cite{Schneier:AppliedCryptography96}. To further
enhance the modulation-based method, the modulating operation
should be non-invertible, i.e., expressing
$s(\mathbf{x},t)=F(g_0(t),\mathbf{x})$, the function
$F(\cdot,\cdot)$ should be non-invertible with respect to
$\mathbf{x}$. However, with such a complicated modulating
mechanism, the synchronization between the sender and the receiver
may become much more difficult and be impossible under practical
conditions (see the next subsection for further discussion on
synchronization issues).

\subsection{The synchronization performance}
\label{section:performance}

In \cite{WuHuZhang:CSF2004}, it was found that ``almost no value
can achieve the synchronization when $M=0$", i.e., when the
original Bu-Wang scheme \cite{BuWang:CSF2004} is used. Our
experiments support this claim. Furthermore, when
$g_0(t)=A\cos(\omega t+\phi_0)$, Matlab's numerical
differentiation function, \texttt{ode45} used in this work, fails
(i.e., enters into a dead lock) with a high probability for many
values of $c$. This implies that the original synchronization
scheme proposed in \cite{BuWang:CSF2004} is ill-behaved in most
cases.

\begin{figure}
\centering\setlength\figwidth{0.45\textwidth}
\includegraphics[width=\textwidth]{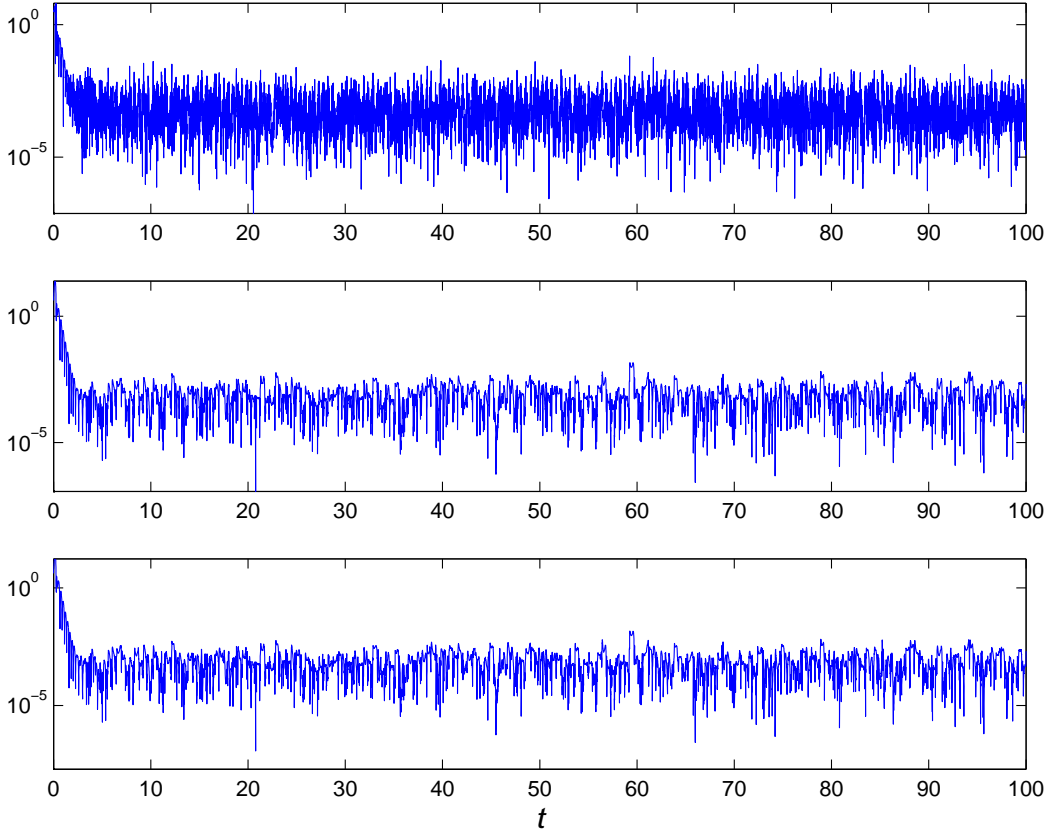}
\caption{The synchronization errors of the three system variables
when $g_0(t)=\mathrm{rand}(t)\in(0,1)$: $|y_1(t)-x_1(t)|$,
$|y_2(t)-x_2(t)|$ and $|y_3(t)-x_3(t)|$ (from top to
bottom).}\label{figure:Syn_noise}
\end{figure}

It can be confirmed that chaos synchronization of the modified
scheme suggested in \cite{WuHuZhang:CSF2004} can be achieved for a
large range of values of $c$. This means that the modification in
\cite{WuHuZhang:CSF2004} really enhances the synchronization
performance. For verification, a number of experiments were
carried out to study the synchronization performance when $g_0(t)$
is replaced by some other functions. As a surprising result, it
was found that even a non-negative noise signal can make the
receiver system synchronized well with the sender system. Figure
\ref{figure:Syn_noise} shows the experimental result when
$g_0(t)=\mathrm{rand}(t)\in(0,1)$ and $c=10$. Some other signals
were also tested, for example, $g_0(t)=t^2\bmod 1$ and
$g_0(t)=(t\cos(\omega t+\phi_0))\bmod 1$, and it was found that
chaos synchronization is achieved in both cases. At present, the
reason behind this unexpected synchronization phenomenon is still
not clear.

\section{Conclusion}

In \cite{BuWang:CSF2004}, a simple modulation-based method was
proposed to improve the security of chaos-based secure
communications against return-map-based attacks. Soon this scheme
was independently cryptanalyzed in \cite{CheeXuBishop:CSF2004,
WuHuZhang:CSF2004, Alvarez:CSF2004} via different attacks. Then,
in \cite{WuHuZhang:CSF2004}, a modified modulation-based scheme
was proposed to enhance the security of the original one. This
paper proposes a new attack to break the new modified scheme.
Compared with previous attacks, the proposed attack is more
powerful and can also break the original scheme in
\cite{BuWang:CSF2004}. Based on all analysis and experimental
results, the conclusion is that, from a cryptographical point of
view, the security of this class of modulation-based scheme is not
satisfactory for secure communications.

\ack{This work was supported by the Applied R\&D Center, City
University of Hong Kong, Hong Kong SAR, China, under Grants no.
9410011 and no. 9620004, and by the Ministerio de Ciencia y
Tecnolog\'{\i}a of Spain, research grant TIC2001-0586 and
SEG2004-02418.}

\bibliographystyle{elsart-num}
\bibliography{CHAOS3330}

\end{document}